\documentclass[a4paper,11pt]{article}
\pdfoutput=1 

\usepackage{jcappub} 

\usepackage[T1]{fontenc} 

\usepackage{braket}
\usepackage{adjustbox}
\usepackage{graphicx,epsfig}
\usepackage{amsmath}
\usepackage {amssymb}
\usepackage {longtable}
\usepackage{multirow}
 \usepackage{mathrsfs}

\usepackage{dcolumn}
\usepackage{bm}
\usepackage{amsfonts}
\usepackage{subfigure}
\usepackage{color}
\usepackage{relsize}

\newcommand{\be}{\begin{equation}}
\newcommand{\ee}{\end{equation}}
\newcommand{\bea}{\begin{eqnarray}}
\newcommand{\eea}{\end{eqnarray}}

\DeclareMathOperator\arctanh{arctanh}

\title{\boldmath Late-time acceleration and structure formation in \break interacting $\alpha$-attractor dark energy models}

\author[a]{L. K. Duchaniya,}
\author[b]{B. Mishra,}
\author[c]{G. Otalora,\note{Corresponding author.}}
\author[d]{M. Gonzalez-Espinoza}

\affiliation[a]{Department of Mathematics, School of Computer Science and Artificial Intelligence, SR University, Warangal 506371, Telangana, India}
\affiliation[b]{Department of Mathematics, Birla Institute of Technology and Science, Pilani, Hyderabad Campus, Jawahar Nagar, Kapra Mandal, Medchal District, Telangana 500078, India.}
\affiliation[c]{Departamento de F\'isica, Facultad de Ciencias, Universidad de Tarapac\'a, Casilla 7-D, Arica, Chile}
\affiliation[d]{Instituto de F\'{\i}sica, Pontificia Universidad Cat\'olica de 
Valpara\'{\i}so, 
Casilla 4950, Valpara\'{\i}so, Chile}

\emailAdd{lokeshkumar@sru.edu.in}
\emailAdd{bivu@hyderabad.bits-pilani.ac.in}
\emailAdd{giovanni.otalora@academicos.uta.cl}
\emailAdd{manuel.gonzalez@pucv.cl}

\abstract{We investigate the cosmological dynamics of interacting dark energy within the framework of $\alpha$-attractor models. Specifically, we analyze the associated autonomous system, focusing on its fixed points that represent dark energy and scaling solutions, along with their stability conditions. We employ Centre manifold theory to address cases where some fixed points display eigenvalues with zero and negative real parts. The model reveals attractors describing dark energy, enabling a smooth transition from the radiation-dominated era to the matter-dominated era, and ultimately into the dark-energy-dominated phase. Additionally, we identify a scaling matter solution capable of modifying the growth rate of matter perturbations during the matter-dominated epoch. Consequently, we study the evolution of matter
perturbations by obtaining both analytical and numerical solutions to the density contrast evolution equation. Based on these results, we compute numerical solutions for the weighted growth rate $f\sigma_{8}$, indicating that interacting $\alpha$-attractor dark energy models may provide a better fit to structure formation data than the standard $\Lambda$CDM scenario.}

\begin{document}

\maketitle
\flushbottom

\section{Introduction}\label{Introduction}
Numerous observational evidence \cite{SupernovaSearchTeam:1998fmf, SupernovaCosmologyProject:1998vns, Spergel_2003WMAP, Tegmark_2004SDSS} from diverse sources indicate that the Universe transitioned into an accelerated expansion phase. The most straightforward interpretation of the observed accelerated expansion involves the incorporation of a positive cosmological constant ($\Lambda$), leading to the standard $\Lambda$CDM model. However, this approach encounters the challenges associated with the cosmological constant problem. Furthermore, the $\Lambda$CDM model exhibits notable phenomenological tensions, most prominently in the measurements of the Hubble constant $H_0$ \cite{Vagnozzi_2020dd, DiValentino:2020zio, Freedman_2021apj} and the matter fluctuation amplitude $\sigma_8$ \cite{Di_Valentino_2021S8} (see also Ref.~\cite{DiValentino:2025sru}). The measurements of the Hubble constant $H_{0}$ reveal a significant tension, currently at the level of around $5\sigma$, between early-Universe estimates based on Cosmic Microwave Background (CMB) observations by the Planck satellite ($H_0 = 67.27 \pm 0.60 \, \text{km s}^{-1} \, \text{Mpc}^{-1}$) \cite{Planck2018} and late-Universe, local measurements using Cepheid-calibrated Type Ia supernovae ($H_0 = 73.04 \pm 1.04 \, \text{km s}^{-1} \, \text{Mpc}^{-1}$) \cite{Riess_2022apj}. This discrepancy is the $H_{0}$ tension. The $\sigma_8$ tension refers to a disparity at the level of approximately $3\sigma$ in the parameter that quantifies the amplitude of matter clustering within spheres of radius $8\, \text{h}^{-1} \, \text{Mpc}$. Specifically, the $\sigma_8$ values inferred from the CMB observations \cite{Planck2018} differ from those obtained through SDSS/BOSS measurements \cite{Alam_2017sdss_III, Ata_2017BAO}. These are just a few problems the concordance cosmological model needs to tackle.  

Two primary approaches are used to address the challenges that have been discussed previously. The first approach preserves the core framework of general relativity (GR) while introducing dynamic components, like a time-dependent cosmological constant or scalar fields, to account for dark energy and the accelerated Universe expansion at late times \cite{Amendola_2000cou, Copeland:2006wr, Tsujikawa_2013Quin}. The second approach is to develop innovative models that demonstrate more complex dynamics at cosmological scales by revising the foundational framework of gravitational theory \cite{DeFelice:2010aj, Capozziello:2011et, Cai:2015emx,Otalora:2022ehd,Burton-Villalobos:2025tgc}. Scalar fields play a crucial role in the standard model of particle physics and have profound implications for a range of cosmological processes. A prime example is the Higgs field, which provides mass to elementary particles through electroweak symmetry breaking. Additionally, other scalar fields are implicated in generating quantum fluctuations during the early Universe, which give rise to the large-scale structure we observe today, as well as to the background of primordial gravitational waves \cite{Kallosh_2013JCAP, Kallosh:2013yoa, Galante_2015PRd, Rodrigues_2021,Gonzalez-Espinoza:2020azh,Gonzalez-Espinoza:2021qnv,Lopez:2021agu,Leyva:2022zhz,ElBourakadi:2024pqr}. Moreover, scalar fields are proposed as potential candidates for dark energy, driving the observed accelerated expansion of the Universe in its later stages.

In this context, $\alpha$-attractor models have emerged as a class of inflationary theories extending into the dark energy sector, particularly within the quintessence model \cite{Linder:2015qxa, Dimopoulos_2017}. Here, the scalar field responsible for dark energy evolves according to a potential shaped by the $\alpha$-attractor mechanism. The $\alpha$ parameter in these models controls the curvature of the potential, which directly influences the dynamics of the scalar field and the evolution of the Universe \cite{Garcia-Garcia:2018hlc, Germ_n_2021, Bhattacharya_2023}. These models are especially noteworthy because they naturally lead to a nearly flat potential, enabling a slow roll of the scalar field that can drive the inflation in the early Universe and the late-time cosmic acceleration \cite{Dimopoulos_2017, Rodrigues_2021,Akrami_2018}. Thus, the $\alpha$-attractor quintessence model provides a unified framework that connects early Universe inflation with the current accelerated expansion, making it a compelling candidate for explaining the nature of dark energy \cite{Kallosh2015plank, Akrami_2018, Shahalam_2018obs, Bag_2018tracker, Mishra_2017alpha}.

The motivation to explore the $\alpha$-attractor class of potentials is that they represent a well-founded category of inflationary potentials, capable of producing universal forecasts for large-scale observables mostly unaffected by the specifics of the inflationary potential. Additionally, these potentials yield predictions that closely align with the current observational constraints on the primordial power spectra \cite{Akrami:2018odb, Plank2020_aghanim}. It is worth noting that the Starobinsky potential \cite{STAROBINSKY198099}, which has been shown to align well with current observations, is a specific case of the $\alpha$-attractor potential.

This article uses the dynamical system analysis to study the dynamics of the $\alpha$-attractor model in the dark energy scenario. To frame our study, we consider the generalized $\alpha$-attractor potential function \cite{Linder:2015qxa}. Complex equations usually characterize cosmological models, and the complexity increases when moving from the background to the perturbation level. As a result, suitable mathematical approaches are crucial for obtaining an analytical understanding that does not depend on initial conditions or specific cosmic development. An effective mathematical instrument for this purpose is dynamic systems analysis. By employing dynamical system analysis, we can investigate the cosmological evolution of the Universe by examining the fixed points of the autonomous system. This approach offers a more profound understanding of the stability and dynamics of cosmological models throughout their evolution. A comprehensive cosmological model can yield various potential outcomes, yet its long-term behavior typically converges toward a stable attractor. This stability is defined by critical points that arise from the autonomous systems governed by the cosmological equations. Fixed points indicate transitional phases in cosmological evolution, which must be unstable nodes or saddle points \cite{Copeland:2006wr, Bahamonde:2017ize}. Numerous researchers have applied dynamical systems theory to extensively analyze cosmological models, investigating their behaviors and properties \cite{Xu:2012jf, Otalora:2013tba, Gonzalez-Espinoza:2020jss, Paliathanasis:2021nqa, Kadam:2022lgq, Gonzalez_Espinoza_2024pha, Duchaniya_2024cqg}. 

The interaction between dark matter and energy is gaining increasing attention as a potential solution to the cosmological coincidence problem. Research in this area has largely centered on a core set of phenomenological interactions, revealing significant implications for the dynamic evolution of the Universe \cite{Andrew:2000prd, Amendola_2000cou, Wang_2016dmde, Amendola_2020jcap, khyllep:2022prd, Rodriguez_Benites_2024epjc}. Recent cosmological observational datasets, provide strong constraints on such interacting scenarios \cite{Silva_2025, Yang:2025, Giar__2024, van_der_Westhuizen_2024, Benisty_2024,Wang:2024vmw}. Given the substantial impact of dark energy and dark matter interactions on the evolution of the Universe and the formation of cosmic structures, conducting a comprehensive dynamical system analysis of interacting cosmology at both the background and perturbation levels is interesting. This article explores the interaction between dark matter and energy by analyzing the system's dynamical behavior and calculating the fixed points and their stability within the relevant cosmological model. Specifically, it examines how changes in the interaction strength affect the evolution of matter perturbations within the large-scale structure of the Universe.

Motivated by this objective, the present study conducts a dynamical system analysis of the $\alpha$-attractor potential within an interacting scalar field model. It further investigates the evolution of matter perturbations for various values of the interaction term.
The structure of this article is organized as follows: In Sec.- \ref{mathformalism}, the mathematical formalism of the model has been presented. The phase space analysis and the behavior of the critical points with the numerical results have been performed in Sec.-\ref{phasespace}. The Centre manifold theory has been incorporated in Sec.-\ref{CMT} and the cosmological perturbation in Sec.-\ref{perturbation}. Finally, the discussions and final remarks are given in Sec.-\ref{conclusions}.  
\section{\texorpdfstring{$\alpha$}{alpha}-attractors and dark energy} \label{mathformalism}
The action of the $\alpha$-attractor model can be defined as \cite{Galante_2015attractor, Linder:2015qxa,Garcia-Garcia:2018hlc},
{\small
\bea \label{alpha_attractor}
 S=\int{d^4 x \sqrt{-g}\left[\frac{R}{2 \kappa^2} -\frac{\alpha\left( \partial{\varphi}\right)^2}{2\left(1-\varphi^2/6\right)}-\alpha f^2\left(\frac{\varphi}{\sqrt{6}}\right) \right]}+S_{m}+S_{r}\,, 
\eea} 

where $(\partial \varphi)^2 = g^{\mu \nu} \partial_{\mu} \varphi \, \partial_{\nu} \varphi$. The terms $S_{m}$ and $S_{r}$ denote the actions for matter and radiation, respectively. The parameter $\alpha$ characterizes the model and $\alpha f^2$ represents the potential function. By performing a field redefinition, $\phi = \sqrt{6\alpha} \, \arctanh\left( \frac{\varphi}{\sqrt{6}} \right)$, the kinetic term of the scalar field takes a canonical form. Consequently, the action can be rewritten as
{\small
\bea \label{alphaatractor_vX}
S=\int{d^4 x \sqrt{-g}\left[\frac{R}{2 \kappa^2} -\frac{\left( \partial{\phi}\right)^2}{2}-V(x)\right]}+
S_{m}+S_{r}\,,
\eea} 

where $x=\tanh(\phi/\sqrt{6 \alpha})$ and $V(x)=\alpha f^2\left(x\right)$. Now, the field space is enlarged within the connected area of $\varphi$ because this transformation extends the boundaries of the initial field, $\varphi \in (-\sqrt{6}, \sqrt{6})$, towards $\pm\infty$ in the transformed field, $\phi \in (-\infty, \infty)$.

We intend to study the generalized $\alpha$-attractor potential \cite{Linder:2015qxa}
\be
V(x)=\alpha c^2\frac{x^p}{\left(1+x\right)^{2n}}=\alpha c^2 2^{-2n} (1-y)^p(1+y)^{2n-p}, 
\label{spot}
\ee 

where $c$, $p$ and $n$ are constant parameters and $y=e^{-2\kappa \phi/\sqrt{6\alpha}}$. We have redefined $\alpha\rightarrow \alpha/\kappa^2$. In cosmological modeling, the specific values of the parameters $\alpha = 1$, $n = 1$ and $p = 2$ are associated with the Starobinsky inflation model \cite{WHITT1984176}. 

In choosing the cosmological background, we assume flat Friedmann-Lema\^{i}tre-Robertson-Walker (FLRW) metric
\begin{equation}
ds^2=-dt^2+a(t)^2\, [dx^2+dy^2+dz^2] \,,
\label{FRWMetric}
\end{equation} 

where $a(t)$ is the scale factor. Hence, the background equations are given as,
\bea
\label{H00}
&& H^2=\frac{\kappa^2}{3}\left[\frac{1}{2}\dot{\phi}^2+V(\phi)+\rho_{m}+\rho_{r}\right],\\
&& \dot{H}=-\frac{\kappa^2}{2}\left[\dot{\phi}^2+\rho_{m}+\frac{4}{3}\rho_{r}\right],\label{Hii}\\
&&\ddot{\phi}+3H\dot{\phi}+\frac{dV}{d\phi}=- \dfrac{Q}{\dot{\phi}},
\label{phiEq}
\eea where $H=\frac{\dot{a}}{a}$ is the Hubble parameter. An overdot indicates the derivative with respect to cosmic time $t$. The interaction term is chosen as
\begin{equation}
Q = \beta \kappa\rho_m\dot{\phi},
\label{Eq1_0}
\end{equation}
originally introduced in \cite{Amendola:1999er}. For $Q>0$, energy is transferred from the dark matter sector to the dark energy sector, while the flow reverses when $Q<0$.
This interaction can be written in the covariant form
\begin{equation}
\nabla_{\mu}T^{\mu}{}_{\nu (\phi)} = C\,T_{(m)}\,\nabla_{\nu}\phi,  
\qquad  
\nabla_{\mu}T^{\mu}{}_{\nu (m)} = -C\,T_{(m)}\,\nabla_{\nu}\phi,
\label{Eq1}
\end{equation}
where $T_{(m)} \equiv T^{\mu}{}_{\mu (m)}$ is the trace of the matter energy–momentum tensor.
For non-relativistic matter, $T_{(m)} \simeq -\rho_m$, so Eq.~\eqref{Eq1} follows directly, while for radiation, $T_{(\gamma)} = 0$, implying no coupling to radiation. Importantly, this formulation preserves general covariance since the total energy–momentum tensor remains conserved.
As emphasized in \cite{Amendola:1999er} (see also \cite{Wetterich:1994bg,Damour:1994ya}), this form arises naturally in scalar–tensor theories—string-inspired dilaton models, Brans–Dicke gravity, and supergravity—after transformation to the Einstein frame. In this formulation, the coupling constant is $C=\beta\kappa$, making $\beta$ dimensionless. The interaction scale is therefore fixed entirely by the Planck mass, introducing no additional arbitrary parameters. This is what we refer to as a minimal coupling, as the interaction strength is determined solely by gravitational physics.
From a phenomenological perspective, this form is also appealing: it is proportional to $\dot{\phi}$, so the interaction is dynamically suppressed when the field is nearly frozen (as in the early Universe), thereby avoiding early-time instabilities and tight CMB constraints, while still allowing potentially observable late-time effects on cosmic expansion and structure growth. While more general possibilities exist—such as non-linear couplings, species-dependent interactions, or disformal terms \cite{Damour:1990tw,Casas:1991ky,Koivisto:2012za}—here we focus on the simplest and most widely studied case \eqref{Eq1_0}, which has both a clear high-energy theoretical origin and well-understood cosmological phenomenology.
Within the $\alpha$-attractor dark energy framework, the slow-roll evolution on the potential plateau naturally suppresses the coupling at high redshift, preserving the attractor dynamics, while allowing mild but potentially measurable deviations at late times.

Interacting dark energy scenarios can, in principle, give rise to effective fifth forces, either through couplings confined to the dark sector or via interactions with baryonic matter \cite{Wang_2016dmde,amendola2010dark}. In the framework considered here, the coupling is strictly limited to the exchange of energy-momentum between dark matter and dark energy, while baryons and radiation evolve independently. This construction prevents any direct coupling to visible matter and thereby excludes long-range fifth forces in the baryonic sector, ensuring compatibility with solar-system and local gravity constraints \cite{Clifton:2011jh}. Within the dark sector, the interaction can mimic an additional attractive force between dark matter particles; however, its strength is tightly constrained by cosmological observations--including CMB anisotropies, BAO measurements, Type Ia supernovae, and weak-lensing surveys. These bounds confine the coupling to small values, guaranteeing that 
any fifth-force--like effects remain consistent with the observed growth of large-scale structure \cite{Amendola_2000cou} .

Recent analyses of coupled dark energy models explicitly interpret the interaction as an effective fifth force acting on dark matter particles while setting the baryonic coupling 
to zero in order to satisfy local gravity tests. Using Planck 2018, BAO, supernovae, cosmic chronometers, and RSD data, it has been shown that the coupling strength is tightly constrained to $|\beta|\lesssim 10^{-2}$. This guarantees that any fifth-force–like effects remain small enough to be consistent with large-scale structure formation while still allowing the interaction to play a role in alleviating cosmological tensions such as the $H_0$ discrepancy \cite{Gomez-Valent:2020mqn,Gonzalez-Espinoza:2023qba}.

According to Ref.\cite{Copeland:2006wr}, the Friedmann Eqs.~(\ref{H00}, \ref{Hii}) can be expressed in conventional form as
\begin{eqnarray}
    \frac{3}{\kappa^{2}}H^{2}&=&\rho_{m}+\rho_{r}+\rho_{de}\,, \label{Friedmann00}\\
    -\frac{2}{\kappa^{2}}\dot{H}&=&\rho_{m}+\frac{4}{3}\rho_{r}+\rho_{de}+p_{de}\,. \label{Friedmann11}
\end{eqnarray}
Therefore, the dark energy sector energy and pressure can be characterized as
\begin{eqnarray}
    \rho_{de}&=&\frac{\dot{\phi}^2}{2}+ V(\phi)\,, \label{rhode}\\
    p_{de}&=&\frac{\dot{\phi}^2}{2}- V(\phi)\,. \label{pde}
\end{eqnarray}
In interacting models, the total energy density of the dark sector is conserved, but the dark energy and dark matter densities evolve as
\begin{eqnarray}
 \dot{\rho}_{de} + 3H (\rho_{de}+p_{de})=-Q \ \label{conservationdarkenergy}\,, \\
  \dot{\rho}_{m} + 3H \rho_{m}= Q \,, \label{matterconservation}\\
  \dot{\rho}_r + 4H \rho_{r}= 0 \label{radiationconservation}\,.
\end{eqnarray}
Furthermore, the dark energy equation-of-state (EoS) parameter can be defined as
\begin{equation}\label{EoSde}
    \omega_{de}=\frac{p_{de}}{\rho_{de}}\,.
\end{equation}
The total EoS parameter can be expressed as
\begin{equation}\label{EosTOTAL}
    \omega_{tot}=\frac{p_{de}+p_r}{\rho_{de}+\rho_{m}+\rho_{r}} \,,
\end{equation}
the deceleration parameter $q$ is connected through
\begin{equation}\label{decelerationpara}
    q=\frac{1}{2}(1+3\omega_{tot})\,.
\end{equation}

In this context, the Universe exhibits accelerated expansion when the deceleration parameter $q<0$ and can also be expressed in terms of the total EoS parameter as $\omega_{tot}<-\frac{1}{3}$.

Finally, we can introduce a key set of cosmological parameters known as the standard density parameters as,
\begin{eqnarray}\label{densityparameters}
\Omega_{m}=\frac{\kappa^2 \rho_m}{3 H^2} \,,\quad \Omega_{r}=\frac{\kappa^2 \rho_r}{3 H^2}\,, \quad \Omega_{de}=\frac{\kappa^2 \rho_{de}}{3 H^2},
\end{eqnarray}
that fulfills the constraint equation
\begin{equation}\label{constaintequation}
    \Omega_{m}+\Omega_{r}+\Omega_{de}=1\,,
\end{equation}
where $\Omega_{m}$, $\Omega_{r}$ and $\Omega_{de}$ are the density parameters for matter, radiation and dark energy phase respectively. Eq.~\eqref{constaintequation} limits the energy density of every component of the Universe, similar to the Friedmann Eq.~\eqref{Friedmann00}; however, it is presented in terms of the density parameters. 

The next section presents a comprehensive dynamical system analysis of the model, with dynamical variables defined based on Eqs.~(\ref{H00}, \ref{Hii}). By selecting appropriate dynamical variables, we shall construct the autonomous system and examine the fixed points, which elucidate the behavior of the model.

\section{Phase-space analysis}\label{phasespace}

To study the dynamics of the model, we introduce the following dimensionless variables
\bea \label{dynamicalvariables}
&& X=\frac{\kappa \dot{\phi}}{\sqrt{6} H},\:\:\:\: Y=\frac{\kappa \sqrt{V}}{\sqrt{3} H}, \:\:\:\: \xi=\frac{\kappa^{2} \rho_{r}}{3 H^{2}},  \nonumber\\  
&& y=e^{-\frac{2\kappa \phi}{\sqrt{6 \alpha}}},\:\:\:\:  \lambda=-\frac{V_{,\phi}}{\kappa V},\:\:\:\: \Gamma=\frac{V V_{,\phi\phi}}{V_{,\phi}^2},
\eea
and the constraint equation
\begin{equation}\label{dynamicalconstraintequation}
 \Omega_{m}+X^2+Y^2+\xi=1 \,.  
\end{equation}

Where $\Omega_{de}=X^2+Y^2$ and $\Omega_{r}=\xi$. By using these dimensionless variables \eqref{dynamicalvariables}, the autonomous system can be expressed as,
\bea
\frac{dX}{dN}&=& \frac{1}{2} \left(3 X^3+\sqrt{6} \beta  X^2+X \left(\xi -3 Y^2-3\right) \right.\nonumber\\
&& \left. +\sqrt{6} \left(\beta  \left(\xi +Y^2-1\right)+\lambda  Y^2\right)\right)\,,\nonumber\\
\frac{dY}{dN}&=& \frac{1}{2} Y \left(\xi +3 X^2-\sqrt{6} \lambda  X-3 Y^2+3\right)\,,\nonumber\\
 \frac{d\xi}{dN}&=& \xi  \left(\xi +3 X^2-3 Y^2-1\right)\,,\nonumber\\
 \frac{d\lambda}{dN}&=&\frac{\sqrt{\frac{3}{2}} \lambda ^2 X \left(y^2 (p-n)+2 n y-n+p\right)}{y (n y-n+p)^2}\,,\nonumber\\
 \frac{dy}{dN}&=& -\frac{2 X y}{\sqrt{\alpha }}\,.
\label{Auto_Sys}
\eea

Where $N=\log(a)$. The EoS parameters and the deceleration parameter in terms of dynamical variables become
\bea
&& \omega_{de} = \frac{ X^2-Y^2}{ X^2+Y^2},  \nonumber \\
&& \omega_{tot} = \frac{\xi }{3}+X^2-Y^2,  \nonumber \\
 && q= \frac{1}{2} \left(\xi +3 X^2-3 Y^2+1\right).
\eea

Now, we will analyze the critical points of the autonomous system \eqref{Auto_Sys}, focusing on their stability and numerical solutions. Based on the behavior exhibited by these critical points, we will determine their relevance to various phases of the Universe. Additionally, we will examine the influence of the matter and dark energy interaction parameter ($Q$) on this dynamic.

\subsection{Critical Points}
We find the critical points by setting the conditions $\frac{dX}{dN}= \frac{dY}{dN} = \frac{d\xi}{dN} =\frac{d\lambda}{dN} =\frac{dy}{dN}= 0$. We analyze the definition of each dynamical variable \eqref{dynamicalvariables} and determine that the acceptable critical points must fulfill the physical viability conditions: $ Y_c \geq 0 $ and $ \xi_c \geq 0 $. The critical points of the system \eqref{Auto_Sys} are presented in Table~\ref{TABLE-I}, while Table~\ref{TABLE-II} displays the corresponding values of their cosmological parameters. In this subsection and subsequent sections, we present the parameters $\Omega_{de}^{(r)}$ and $\Omega_{de}^{(m)}$, which denote the fractional density of dark energy in the eras dominated by radiation and dark matter respectively.

Critical points $a_R$ and $b_R^{\pm}$ are associated with the radiation phase where $ \Omega_r=1 $ and $\omega_{tot}=\frac{1}{3} $. The positive value of the deceleration parameter indicates that these points represent a decelerating phase of the Universe. The critical point $c_R$ indicates a scaling radiation era characterized by the expressions for density and the EoS parameters: $\Omega^{(r)}_{de} = \frac{1}{6\beta^2}$, $\omega_{de} = 1$, and $\omega_{tot} = \frac{1}{3}$. This regime must adhere to the stringent constraints set by Big Bang Nucleosynthesis (BBN), specifically ensuring that $\Omega^{(r)}_{de} < 0.045 $ remains consistent with the physics of early Universe \cite{Ferreira_1998, Bean:2001wt}.

Critical points $d_M$ and $e^{\pm}_M$ are linked to the matter-dominated epoch characterized by $\Omega_m=1 $ and $ \omega_{tot}=0$. The positive deceleration parameter signifies that these points correspond to a decelerating phase of the Universe. In contrast, at $\beta = 0$ (there is no interaction between the dark matter and dark energy ), the critical point $f$ signifies a matter-dominated epoch characterized by parameters $\Omega_m = 1$, $\omega_{de} = 1$ and $\omega_{tot} = 0$. For non-zero $\beta$, we observe a scaling matter era characterized by the density parameter $\Omega_{de}^{(m)} = \frac{2 \beta^2}{3}$. This parameter must adhere to the constraint $\Omega_{de}^{(m)} < 0.02$ at a 95\% confidence level, specifically at redshift $z \approx 50$, as indicated by recent CMB observations \cite{Planck:2015bue}.

Critical points $g$ and $i^{\pm}$ represent the de Sitter solutions characterized by $\Omega_{de} = 1 $ and the EoS parameter $ \omega_{de} = \omega_{tot} = -1 $. This configuration leads to accelerated expansion across all parameter values.

The point $j^{\pm}$ represents a solution dominated by dark energy, characterized by $\Omega_{de} = 1$. However, this solution fails to account for the observed accelerated expansion of the Universe, as it behaves similarly to stiff matter, with the EoS parameters $ \omega_{de} = \omega_{tot} = 1$.

The critical point \( k \) resembles a late-time dark energy/matter scaling solution since it is characterized by $\Omega_{de} = 1 + \frac{3}{\beta^2} $, a deceleration parameter \( q = -1 \) and EoS parameter given by \( \omega_{\text{tot}} = -1 \). However, the matter density parameter at this point is \( \Omega_m = -\frac{3}{\beta^2} \), which is non-physical since it implies a negative matter density. Therefore, this critical point \( k \) lacks physical viability.

In the next section, we will analyze the stability of the identified critical points. This analysis will be conducted via a linear perturbation approach to the dynamical variables. It is worth emphasizing that the bounds arising from BBN at the radiation scaling point $c_R$ 
and from CMB constraints at the matter scaling point $f$ appear mutually incompatible if these 
solutions were to represent long-lasting cosmological phases. In practice, however, the 
dynamical trajectories are not necessarily attracted to $c_R$. A viable cosmic history can 
proceed through the sequence $a_R \rightarrow f \rightarrow g$, where the Universe evolves from 
a pure radiation era consistent with BBN, passes through a transient scaling matter regime, and 
finally approaches the late-time accelerated attractor. In this context, the illustrative 
parameter values adopted here are chosen to explore the qualitative effects of the interaction 
on background evolution and structure formation, rather than to represent best-fit observational 
constraints.

\begin{table}[ht]
     \renewcommand{\arraystretch}{2.5} 
     \setlength{\tabcolsep}{20pt} 
    \caption{Critical points for the dynamical system. } 
    \centering 
    \begin{adjustbox}{width=\textwidth}
    \begin{tabular}{ccccccc} 
    \hline\hline 
    C.P. & $X_{c}$ & $Y_{c}$ & $\xi_{c}$ & $\lambda_{c}$ & $y_{c}$ & Exists for  \\ [0.5ex] 
    \hline\hline 
    $a_{R}$ &$0$ & $0$ & $1$ & $\lambda_{c}$&$y_{c}$ & Always \\
    $b^{\pm}_{R}$  & 0 & 0 & 1 & $\lambda_{c}$ & $\frac{n\pm\sqrt{2 n p-p^2}}{n-p}$ & $n \neq p, p(2n-p)> 0$\\
    $c_{R}$  & $-\frac{1}{\sqrt{6} \beta }$ & $0$ & $\frac{2 \beta ^2-1}{2 \beta ^2}$ & $0$ & $0$ & Always \\
    $d_{M}$  & $0$ & $0$ & $0$ & $\lambda_{c}$ & $y_{c}$ & Always \\
    $e^{\pm}_{M}$ & 0 & 0 & $0$ & $\lambda_{c}$ & $\frac{n\pm\sqrt{2 n p-p^2}}{n-p}$ & $n \neq p, p(2n-p)> 0$ \\
    $f$ & $-\sqrt{\frac{2}{3}} \beta$ & $0$ & 0 & 0 & $0$ & Always \\
    $g$ & 0 & $1$ & 0 & 0 & $y_{c}$ & Always\\
    $i^{\pm}$  & 0 & 1 & 0 & 0 & $\frac{n\pm\sqrt{2 n p-p^2}}{n-p}$ & $n \neq p, p(2n-p)> 0$\\
    $j^{\pm}$  & $\pm 1$ & 0 & 0 & 0 & 0 & Always\\
    $k$  & $\frac{\sqrt{\frac{3}{2}}}{\beta }$ & $ \sqrt{1 +\frac{3}{2\beta^2 }}$ & 0 & 0 & 0 & Always\\
    [1ex] 
    \hline\hline 
    \end{tabular}
    \end{adjustbox}
    \label{TABLE-I}
\end{table}


\begin{table}[ht]
     \renewcommand{\arraystretch}{2.5} 
     \setlength{\tabcolsep}{25pt} 
    \caption{ Background Cosmological parameters for the critical points in TABLE-- \ref{TABLE-I}{}} 
    \centering 
    \begin{adjustbox}{width=\textwidth}
    \begin{tabular}{ccccccc} 
    \hline\hline 
    C.P. & $\Omega_{de}$ & $\Omega_{m}$ & $\Omega_{r}$ & $\omega_{tot}$ & $\omega_{de}$ & $q$ \\ [0.5ex] 
    \hline\hline 
    $a_{R}$ &$0$ & $0$ & $1$ & $\frac{1}{3}$&$1$ & 1\\
    $b^{\pm}_{R}$  & 0 & 0 & 1 & $\frac{1}{3}$ & $1$& 1\\
    $c_{R}$  & $\frac{1}{6 \beta ^2}$ & $\frac{1}{3 \beta ^2}$ & $1-\frac{1}{2 \beta ^2}$ & $\frac{1}{3}$ & $1$& $1$\\
    $d_{M}$  & $0$ & $1$ & $0$ & $0$ & $1$ &$\frac{1}{2}$  \\
    $e^{\pm}_{M}$ & 0 & 1 & $0$ & $0$ & $1$&$\frac{1}{2}$ \\
    $f$  & $\frac{2 \beta ^2}{3}$ & $1-\frac{2 \beta ^2}{3}$ & 0 & $\frac{2 \beta ^2}{3}$ & $1$& $\frac{1}{2}+\beta^2$\\
    $g$ & 1 & 0 & 0 & -1 & -1& -1\\
    $i^{\pm}$ & 1 & 0 & 0 & -1 & -1& -1\\
    $j^{\pm}$ & 1 & 0 & 0 & 1 & 1& 2\\
    $k$ & $\frac{3}{\beta ^2}+1$ & $-\frac{3}{\beta ^2}$ & 0 & $-1$ & $-\frac{\beta ^2}{\beta ^2+3}$& -1\\
    [1ex] 
    \hline\hline 
    \end{tabular}
    \end{adjustbox}
    \label{TABLE-II}
\end{table}

\subsection{Stability of critical points}\label{Stability}

Now, we introduce small perturbations around each critical point and discuss how these perturbations help to linearize the equations of the system. This leads to the formation of a perturbation matrix $\mathcal{M}$, where the eigenvalues $\mu_1$, $\mu_2$, $\mu_3$, $\mu_4$ and $\mu_5$ play a crucial role in the determination of stability. The classification of stability generally includes (i) a stable node characterized by all negative eigenvalues, (ii) an unstable node identified by all positive eigenvalues, (iii) a saddle point that exhibits mixed eigenvalues, and (iv) a stable spiral that has a negative determinant and negative real parts. Attractor points, such as stable nodes or spirals, are attained during cosmic evolution regardless of initial conditions provided that they reside within the attraction basin. The critical points exhibit negative real parts and zero eigenvalues, called non-hyperbolic critical points. In such situations, linear stability theory is insufficient for assessing stability. Consequently, we have analyzed the stability of the non-hyperbolic critical points using the Centre manifold theory. Following this, the eigenvalues and stability conditions of each critical point are detailed below.

\begin{itemize}

\item Critical point $a_{R}$ has eigenvalues
\be
\mu_{1} = 0, \hspace{0.2cm} \mu_{2} = 0, \hspace{0.2cm} \mu_{3} = -1, \hspace{0.2cm} \mu_{4} = 1, \hspace{0.2cm} \mu_{5} = 2\,. 
\ee
The critical point exhibits saddle point behavior, characterized by its instability due to the presence of both positive and negative eigenvalues.

\item Critical point $b^{\pm}_{R}$ has eigenvalues
\be
\mu_{1} = 0, \hspace{0.2cm} \mu_{2} = 0, \hspace{0.2cm} \mu_{3} = -1, \hspace{0.2cm} \mu_{4} = 1, \hspace{0.2cm} \mu_{5} = 2\,. 
\ee
Similar to critical point $a_{R}$, this critical point also shows the saddle behavior.

\item Critical point $c_{R}$ has eigenvalues
\bea
 \mu_{1} = 0, \hspace{0.2cm} \mu_{2} = \frac{\sqrt{\frac{2}{3}}}{\sqrt{\alpha } \beta }, \hspace{0.2cm} \mu_{3} = 2, \hspace{0.2cm} \mu_{4} = -\frac{\sqrt{  2-3 \beta ^2}+\beta }{2 \beta },
 \hspace{0.2cm} \mu_{5} = \frac{\sqrt{2-3 \beta ^2}}{2 \beta }-\frac{1}{2}\,.
\eea
This critical point also exhibits saddle point behavior for any choice of $\beta$, as shown by the eigenvalues $\mu_4$ and $\mu_5$, which have opposite signs.

\item Critical point $d_{M}$ has eigenvalues
\be
\mu_{1} = 0, \hspace{0.2cm} \mu_{2} = 0, \hspace{0.2cm} \mu_{3} = -\frac{3}{2}, \hspace{0.2cm} \mu_{4} = -1, \hspace{0.2cm} \mu_{5} = \frac{3}{2}\,. 
\ee
This critical also shows the saddle behavior.

\item Critical point $e^{\pm}_{M}$ has eigenvalues
\be
\mu_{1} = 0, \hspace{0.2cm} \mu_{2} = 0, \hspace{0.2cm} \mu_{3} = -\frac{3}{2}, \hspace{0.2cm} \mu_{4} = -1, \hspace{0.2cm} \mu_{5} = \frac{3}{2}\,. 
\ee
This critical point also demonstrates the saddle-like behavior.

\item Critical point $f$ has eigenvalues
\bea
\mu_{1} = 0, \hspace{0.2cm} \mu_{2} = \frac{2 \sqrt{\frac{2}{3}} \beta }{\sqrt{\alpha }}, \hspace{0.2cm} \mu_{3} = \beta ^2-\frac{3}{2}, \hspace{0.2cm} \mu_{4} = 2 \beta ^2-1, \hspace{0.2cm} \mu_{5} = \beta ^2+\frac{3}{2}\,. 
\eea
This critical point also illustrates saddle point characteristics for any value of $\beta$, as shown by the eigenvalues $\mu_3$ and $\mu_5$, which exhibit opposite signs.

\item Critical point $g$ has eigenvalues
\be
\mu_{1} = 0, \hspace{0.2cm} \mu_{2} = 0, \hspace{0.2cm} \mu_{3} = -4, \hspace{0.2cm} \mu_{4} = -3, \hspace{0.2cm} \mu_{5} = -3\,. 
\ee
The eigenvalues at this critical point exhibit a negative real part and zero. As a result, this point is classified as a non-hyperbolic critical point. Consequently, we will analyze the stability of this critical point using the Centre manifold theory in the upcoming section.

\item Critical point $i^{\pm}$ has eigenvalues
\be
\mu_{1} = 0, \hspace{0.2cm} \mu_{2} = 0, \hspace{0.2cm} \mu_{3} = -4, \hspace{0.2cm} \mu_{4} = -3, \hspace{0.2cm} \mu_{5} = -3\,. 
\ee
Similar to critical point $g$, this critical point also exhibits non-hyperbolic characteristics. Consequently, the stability of this critical point is also evaluated using the Centre manifold theory.

\item Critical point $j^{\pm}$ has eigenvalues
\be
\mu_{1} = \pm \frac{2}{\sqrt{\alpha }}, \hspace{0.2cm} \mu_{2} = 0, \hspace{0.2cm} \mu_{3} = 2, \hspace{0.2cm} \mu_{4} = 3, \hspace{0.2cm} \mu_{5} = 3\,. 
\ee
The critical point $j^+$ exhibits characteristics of an unstable node, whereas the critical point $j^-$ displays saddle-like characteristics.

\item Critical point $k$ has eigenvalues
{\small
\bea
&& \mu_{1} = 0, \hspace{0.2cm} \mu_{2} = -\frac{\sqrt{6}}{\sqrt{\alpha } \beta }, \hspace{0.2cm} \mu_{3} = -4, \nonumber\\
&& \hspace{0.2cm} \mu_{4} = -\frac{3 \left(\sqrt{5 \beta ^2+6}+\beta \right)}{2 \beta }, \hspace{0.2cm} \mu_{5} = \frac{3}{2} \left(\frac{\sqrt{5 \beta ^2+6}}{\beta }-1\right)\,. \nonumber\\
&&
\eea}
This critical point also demonstrates saddle behavior for any selection of $\beta$.
\end{itemize}

\subsection{Numerical Results}

In this section, we numerically solve the autonomous system given by Eqs.~\eqref{Auto_Sys}, which corresponds to the set of cosmological Eqs.~(\ref{H00}--\ref{phiEq}). Our analysis focuses on the evolution of cosmological parameters and their agreement with observational constraints.

\begin{figure}[!h]
    \centering
        \includegraphics[scale=0.65]{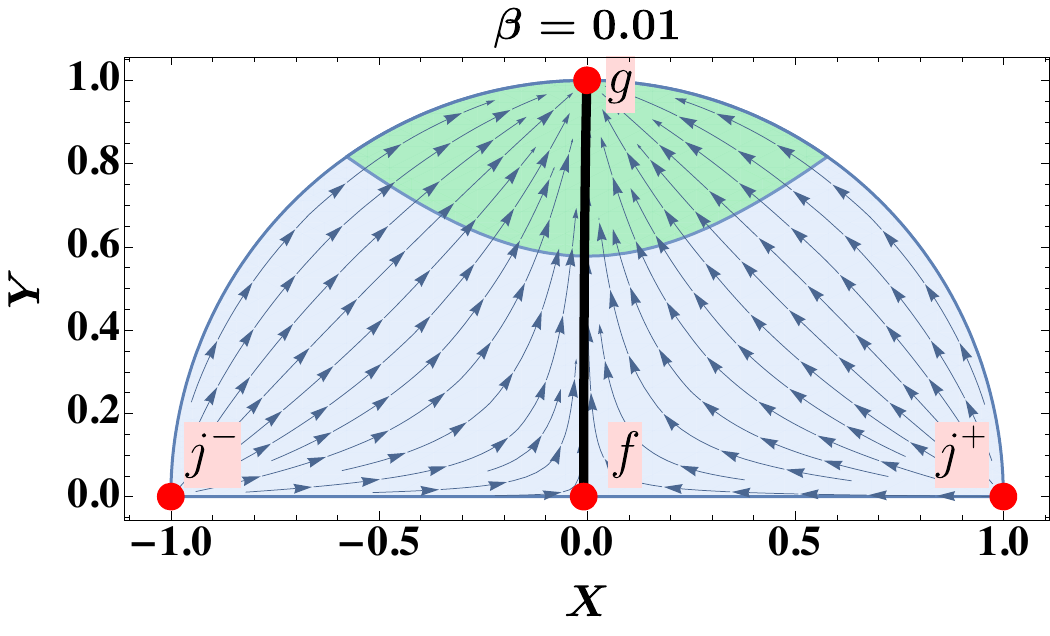}
        \caption{The phase space of $X$ vs. $Y$ for $\beta = 0.01$. We assume $\xi = 0$ and $\lambda = 0$, which are consistent with critical point $g$, to reduce the phase space to two dimensions. The critical points are depicted by red points along with their corresponding labels. Each streamline represents the Universe's evolution under different initial conditions. In particular, the black line presents a Universe with initial conditions $X_0=10^{-11}$, $Y_0=8.6 \times 10^{-13}$, $\xi_0 = 0.999656$, $\lambda_0 = 1.0 \times 10^{-10}$ and  $y_0 = 1.0 \times 10^{-11}$. The green region indicates accelerated expansion.}  
    \label{2Dphase}
\end{figure}

\begin{figure}[!h]
    \centering
        \includegraphics[scale=0.5]{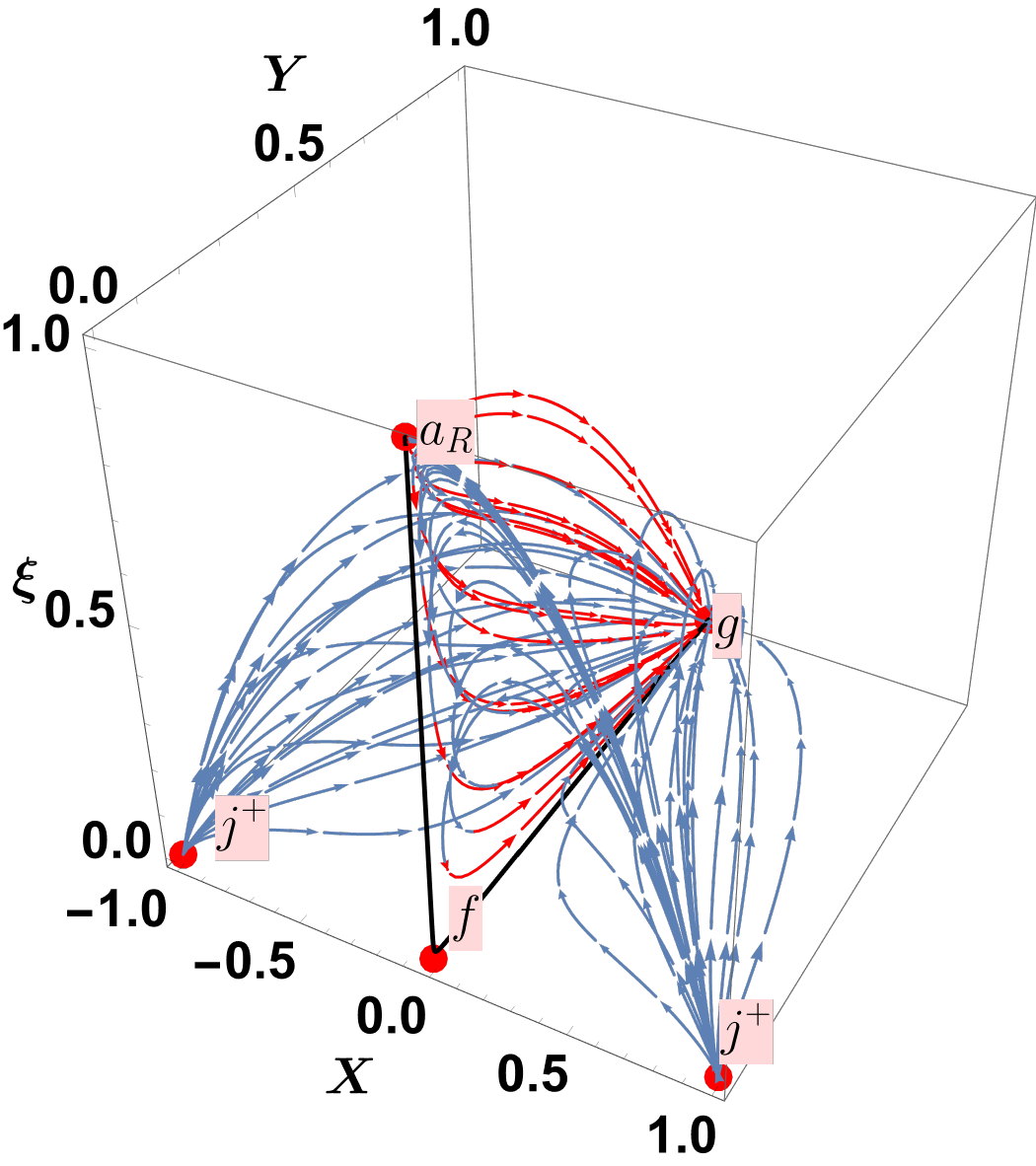}
        \caption{The phase space of $X$ vs. $Y$ vs. $\xi$ for $\beta = 0.01$. We assume $\lambda = 0$, which is consistent with critical point $g$. The critical points are depicted by red points along with their corresponding labels. Each streamline represents the Universe's evolution under different initial conditions. In particular, the black line presents a Universe with initial conditions $X_0=10^{-11}$, $Y_0=8.6 \times 10^{-13}$, $\xi_0 = 0.999656$, $\lambda_0 = 1.0 \times 10^{-10}$ and  $y_0 = 1.0 \times 10^{-11}$.}  
    \label{3Dphase}
\end{figure}

\begin{figure}[!h]
    \centering
        \includegraphics[scale=0.5]{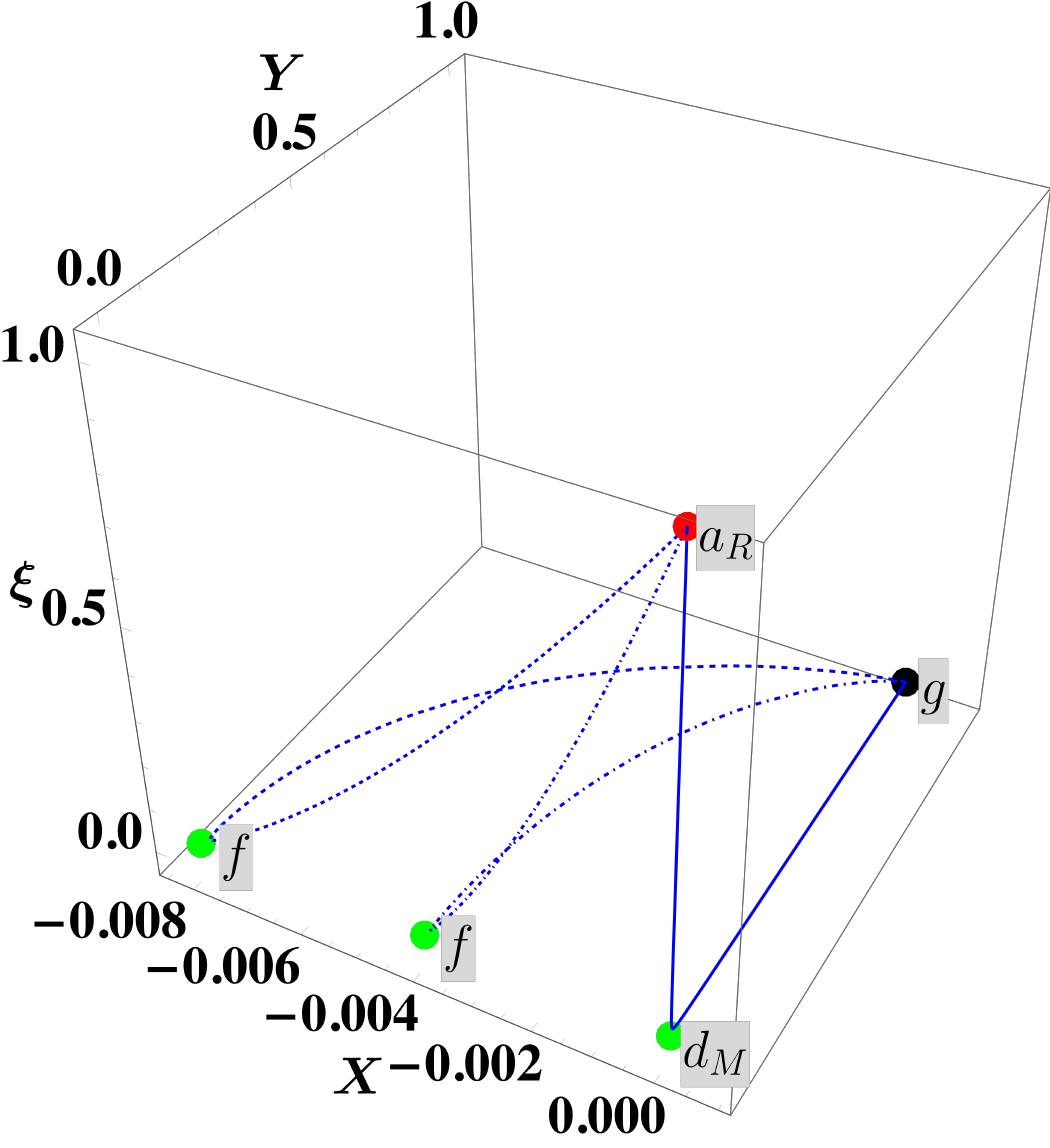}
        \caption{Here, we show phase space for the fixed values $\alpha = 1$, $n=2$ and $p=1$  but varying the values of $\beta$, with $\beta=5.0 \times 10^{-3}$ (dot-dashed line), $\beta=1.0 \times 10^{-2}$ (dash line), and $\beta = 0$ (solid line). And initial conditions $X_0=10^{-11}$, $Y_0=8.6 \times 10^{-13}$, $\xi_0 = 0.999656$, $\lambda_0 = 1.0 \times 10^{-10}$ and  $y_0 = 1.0 \times 10^{-11}$.}  
    \label{numerical_phase}   
\end{figure}

\begin{figure}[!h]
    \centering
        \includegraphics[scale=0.5]{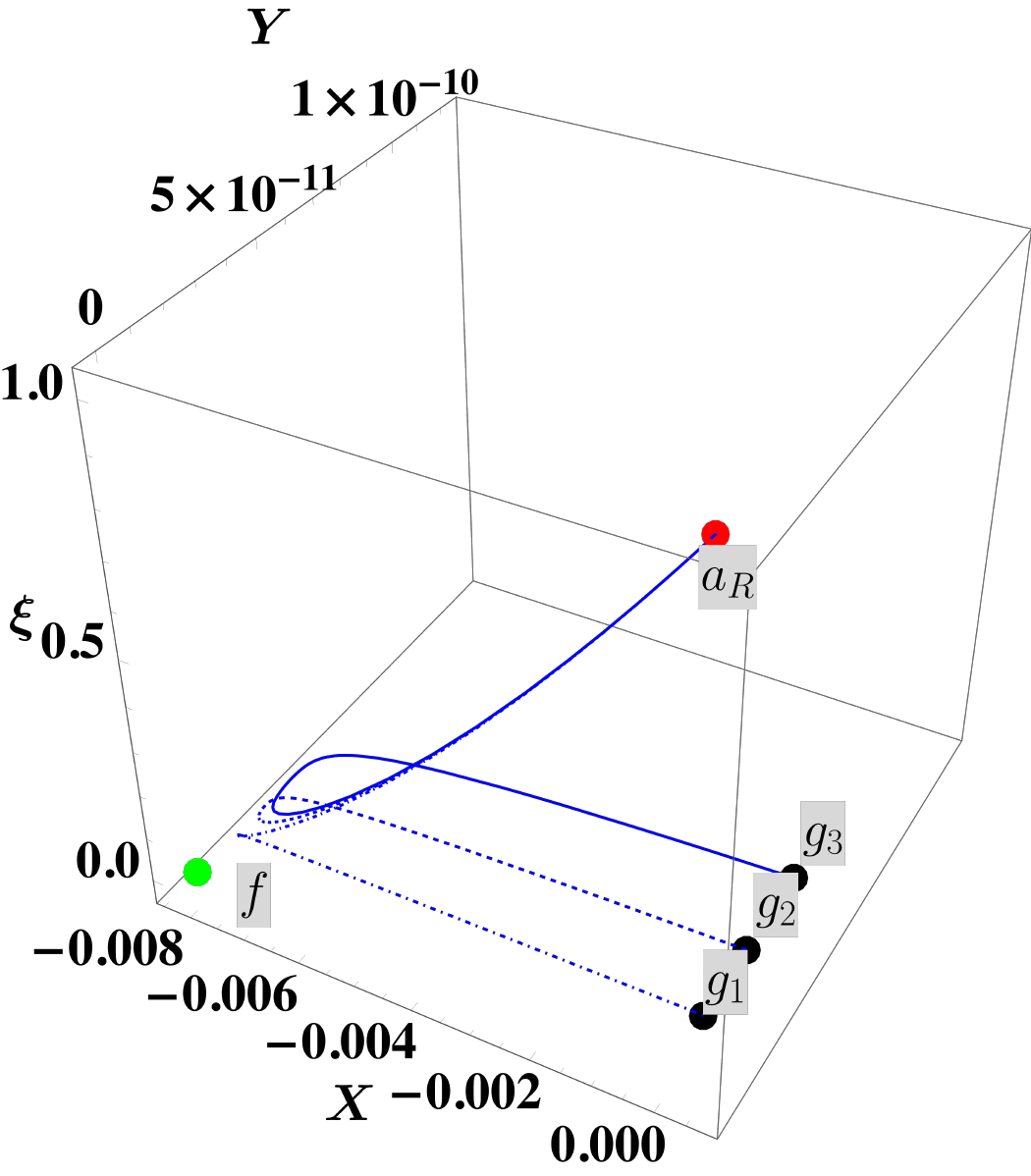}
        \caption{Here, we show phase space for the values $\beta = 1.0 \times 10^{-2}$, $n=2$ and $p=1$  but varying the values of $\alpha$, with $\alpha=1.0 \times 10^{-1}$ (dot-dashed line), $\alpha=1.0 \times 10^{-2}$ (dash line), and $\beta = 5.0 \times 10^{-3}$ (solid line). And initial conditions $X_0=10^{-11}$, $Y_0=8.6 \times 10^{-13}$, $\xi_0 = 0.999656$, $\lambda_0 = 1.0 \times 10^{-10}$ and  $y_0 = 1.0 \times 10^{-11}$.}  
    \label{numerical_phase_alpha}
\end{figure}

\begin{figure}[!h]
    \centering
        \includegraphics[scale=0.6]{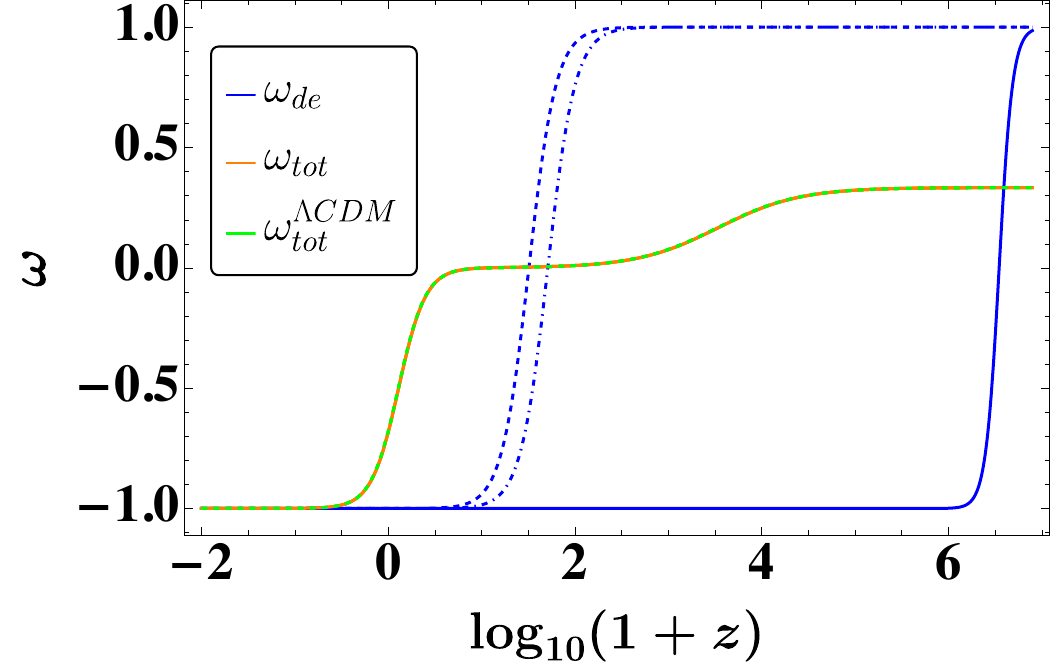}
        \caption{The plot illustrates the evolution of the EoS parameters under the same conditions as those presented in Figure~\ref{numerical_phase}.}
    \label{plotwtotal}
\end{figure}

\begin{figure}[!h]
    \centering
        \includegraphics[scale=0.6]{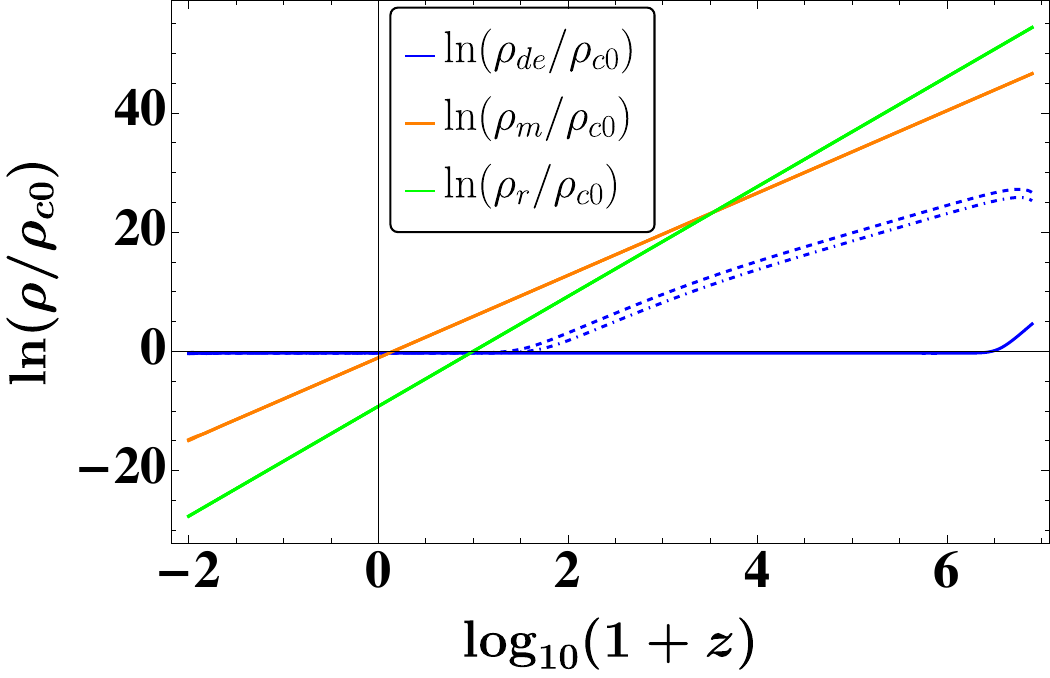}
        \caption{We illustrate the evolution of the energy density of dark energy ($\rho_{\text{de}}$) in blue, dark matter (including baryons) ($\rho_m$) in orange, and radiation ($\rho_r$) in green as functions of the redshift ($z$), for two different values of $\lambda$ and $\alpha$. The conditions considered are the same as those presented in Figure~\ref{numerical_phase}.
}
    \label{plotrhototal}
\end{figure} 
We identify three attractor solutions associated with dark energy-dominated epochs exhibiting cosmic acceleration: points $g$, $i^{+}$, and $i^{-}$. Additionally, we identify the scaling solution $f$, characterizing the scaling matter epoch.

Figures \ref{2Dphase}, \ref{3Dphase}, and \ref{numerical_phase} illustrate phase space trajectories that show transitions such as \( a_R \to d_M \to g \) and \( a_R \to f \to g \). Each trajectory represents different values of \( \beta \). Additionally, Figure \ref{numerical_phase_alpha} shows the phase space trajectories \( a_R \to f \to g \) for varying values of \( \alpha \). Figure \ref{plotwtotal} presents the total EoS parameter $w_{\text{tot}}$ and the dark energy EoS parameter $w_{\text{de}}$, where the latter in $z = 0$ is found to be $w_{\text{de}}(z=0) \approx -1.01$, consistent with the observational constraint $w_{\text{de}}^{(0)} = -1.028 \pm 0.032$ \cite{Planck:2018vyg}. In Figure~\ref{plotrhototal}, we display the evolution of the energy densities of dark energy, dark matter, and radiation, which are consistent with present values of the fractional energy densities as $\Omega_{\text{de}}^{(0)} \approx 0.68$ and $\Omega_{\text{m}}^{(0)} \approx 0.32$.

During the scaling matter regime, for trajectories $a_R \to f \to g$, we impose constraints on the fractional energy density of dark energy, $\Omega_{\text{de}}^{(m)}$, CMB measurements \cite{Planck:2015bue}. Specifically, at point $f$, we find $\Omega_{\text{de}}^{(m)} \approx 1.27 \times 10^{-4}$ at $z = 50$.

\begin{figure}[!t]
    \centering
        \includegraphics[scale=0.7]{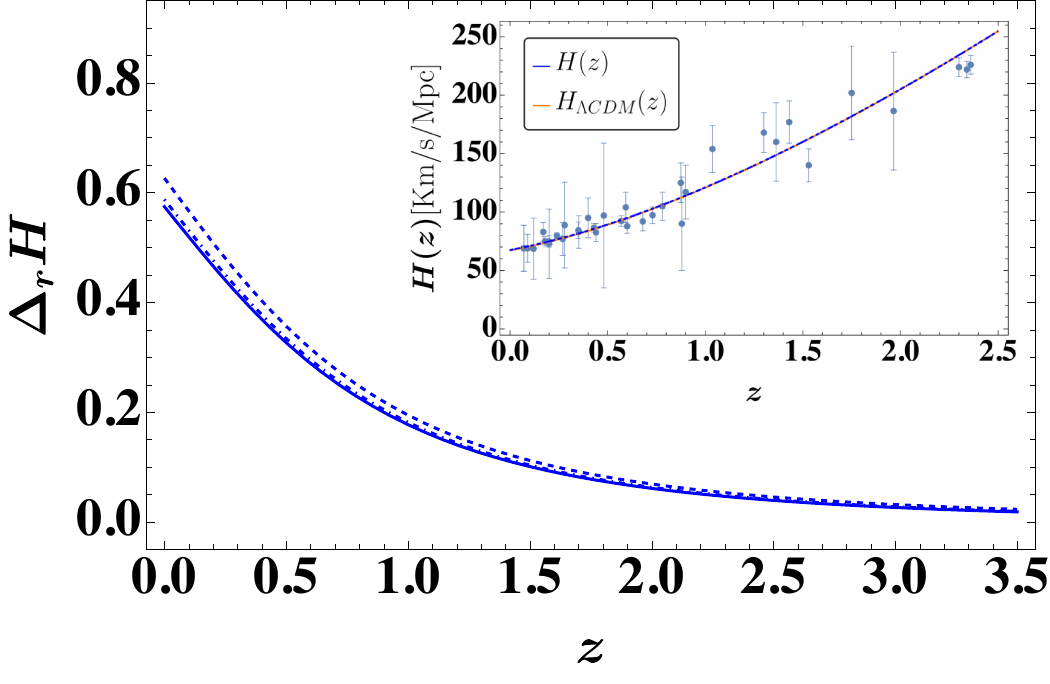}
        \caption{This plot illustrates the evolution of the relative difference $\Delta_r H$ with respect to the $\Lambda$CDM model as a function of the redshift $z$. Additionally, the inset presents a comparison between the Hubble parameter $H(z)$ obtained from our model and the Hubble parameter associated with the $\Lambda$CDM model, denoted as $H_{\Lambda \text{CDM}}(z)$. Observational data points are also included for reference. The conditions considered are the same as those presented in Figure~\ref{numerical_phase}.
}
    \label{plotdH}
\end{figure}

Figure~\ref{plotdH} illustrates the evolution of the Hubble parameter $H(z)$, comparing our model with the $\Lambda$CDM prediction and observational Hubble data \cite{Cao:2021uda, Farooq:2013hq}. The results show that our model closely follows $\Lambda$CDM, satisfying preliminary viability criteria.

\begin{figure}[!t]
    \centering
        \includegraphics[scale=0.7]{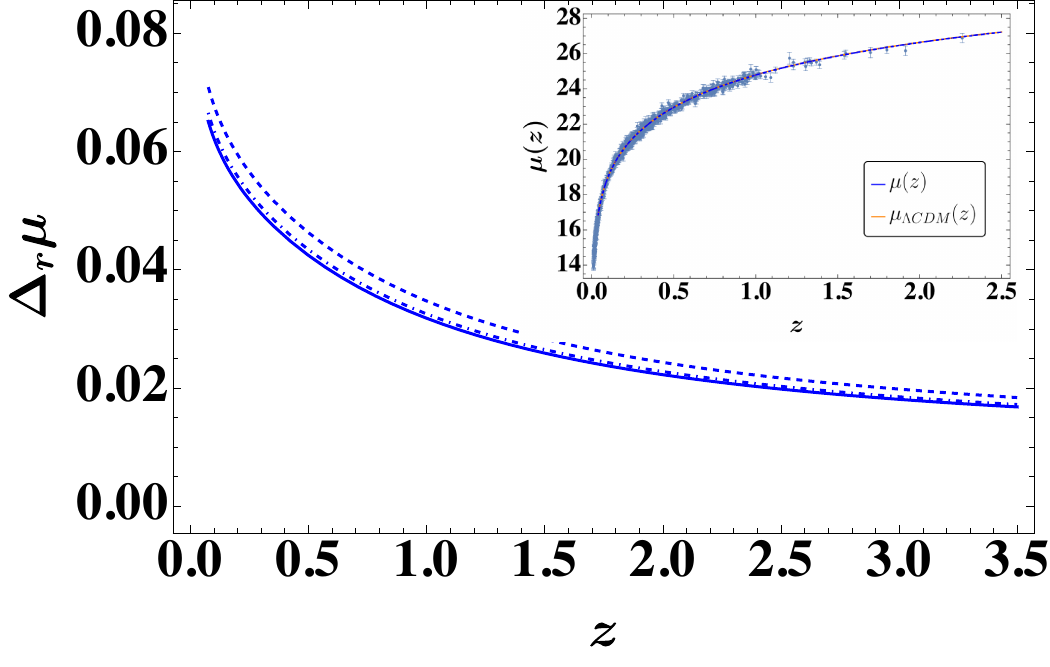}
        \caption{The plot illustrates the evolution of the relative difference $\Delta \mu_{r}$ with respect to the $\Lambda$CDM model as a function of the redshift $z$. Additionally, within this figure, we present the evolution of the distance modulus $\mu(z)$ for our first interacting model, alongside the distance modulus associated with the $\Lambda$CDM model, denoted as $\mu_{\Lambda \text{CDM}}(z)$, as a function of $z$. The conditions considered are the same as those presented in Figure~\ref{numerical_phase}.
}
    \label{plotdDL}
\end{figure}

Figure~\ref{plotdDL} presents the evolution of the distance modulus $\mu(z)$ alongside the corresponding $\Lambda$CDM model prediction, $\mu_{\Lambda \text{CDM}}(z)$, as a function of $z$. We particularly focus on the Type Ia Supernovae data in the Pantheon sample \cite{Pan-STARRS1:2017jku}, which consists of 1048 supernovae spanning the redshift range $0.01 < z < 2.3$.\footnote{Data available online in the GitHub repository: \url{https://github.com/dscolnic/Pantheon}.}

\begin{figure}[!t]
    \centering
        \includegraphics[scale=0.7]{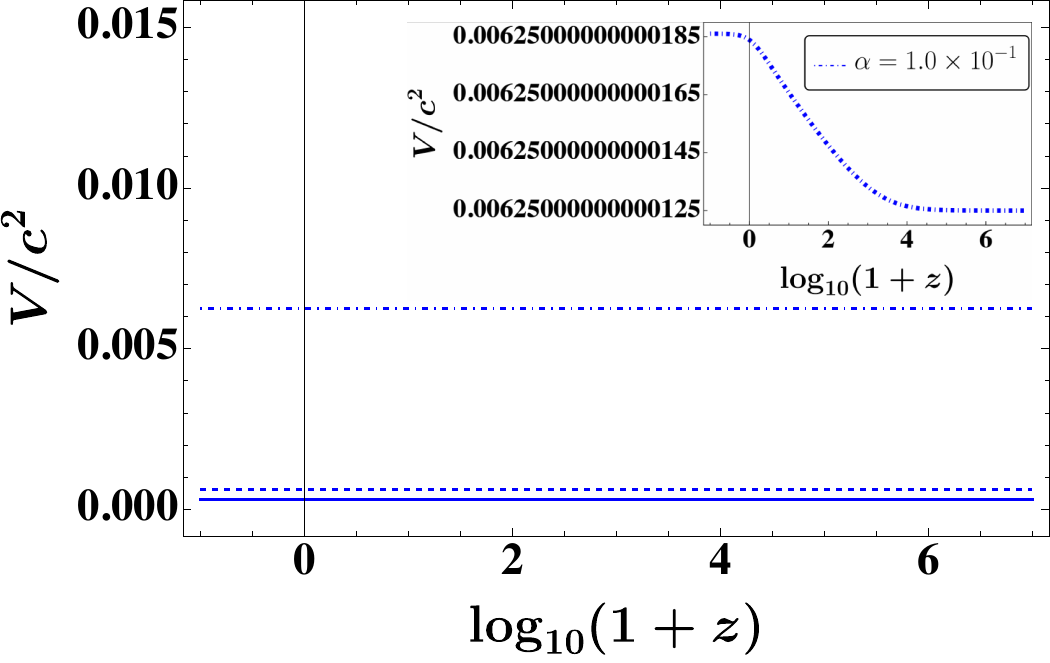}
        \caption{The plot illustrates the evolution of the $\alpha$-attractor potential for the values $\beta = 1.0 \times 10^{-2}$, $n = 2$, and $p = 1$, while varying the values of $\alpha$. Specifically, the lines represent $\alpha = 1.0 \times 10^{-1}$ (dot-dashed line), $\alpha = 1.0 \times 10^{-2}$ (dashed line), and $\beta = 5.0 \times 10^{-3}$ (solid line). The initial conditions are set as follows: $X_0 = 10^{-11}$, $Y_0 = 8.6 \times 10^{-13}$, $\xi_0 = 0.999656$, $\lambda_0 = 1.0 \times 10^{-10}$, and $y_0 = 1.0 \times 10^{-11}$. Additionally, the figure includes a zoomed-in view of the $\alpha$-attractor potential for $\alpha = 1.0 \times 10^{-1}$.
}
    \label{Vpotential}
\end{figure}

Figure \ref{Vpotential} illustrates the evolution of the $\alpha$-attractor potential for varying values of $\alpha$. The results show how the shape of the potential changes with $\alpha$, particularly for small values. This variation in the potential is directly related to the behavior observed in the phase space trajectories shown in Figure \ref{numerical_phase_alpha}, where different values of $\alpha$ influence the transition paths. Nevertheless, it is important to note that the overall cosmological dynamics remain largely independent of $\alpha$.  

In summary, our numerical analysis uncovers new scaling solutions and attractor points within the $\alpha$-attractor framework in dark energy scenarios. These solutions naturally incorporate early dark energy and offer a phenomenological alternative to the $\Lambda$CDM model. Further investigations, including best-fit parameter estimation from observational data, are required to establish the model's full viability.

\section{Centre Manifold theory} \label{CMT}
Centre Manifold Theory (CMT) is a specialized area within dynamical systems theory that focuses on analyzing the local behavior of nonlinear systems near fixed points. Perko \cite{Perko2001} provides a comprehensive exposition of the mathematical foundations of CMT. Traditional linear stability analysis often proves inadequate for determining the stability of critical points when the associated Jacobian matrix possesses eigenvalues with zero real parts. In contrast, CMT enables a meaningful stability analysis by effectively reducing the dimensionality of the system near such critical points. When a system evolves through a critical point, its local dynamics are governed by the invariant Centre manifold, denoted as $W_c$. This manifold is associated with the eigenvalues having zero real parts, and the behavior confined to $W_c$ encapsulates the essential characteristics of the system near fixed points \cite{Bahamonde:2017ize}.

Let us consider the dynamical system governed by the following set of equations
 \bea \label{CMT1}
\tau'= F(\tau)\,,
\eea

where $\tau= (\mu, \nu)$. A geometric space is considered a Centre manifold for this system if it can be expressed locally as,
\bea \label{invariantmanifold}
W_c = \{ (\mu, \nu) \in \mathbb{R} \times \mathbb{R} : \nu = h(\mu), |\mu| < \delta, h(0) = 0, \nabla h(0) = 0 \}\,,
\eea
for sufficiently small $\delta$, $h(\mu)$ is a smooth function on $\mathbb{R}$. 
The following steps outline the detailed analysis of the CMT:

\begin{enumerate}
    \item \textbf{Coordinate shift:} First, the positions of the non-hyperbolic critical points are moved to the origin, leading to a new set of autonomous equations represented in the transformed coordinate system.
    \item \textbf{Reformulation of the Dynamical System:} The modified dynamical system is presented in the standard format, allowing for additional analysis, as expressed below,
    \bea \label{ReformulationoftheDynamicalSystem}
\tau' =
\begin{pmatrix}
\mu' \\ 
\nu'
\end{pmatrix}
=
\begin{pmatrix}
A\mu \\ 
B\nu
\end{pmatrix}
+
\begin{pmatrix}
\phi(\mu, \nu) \\ 
\psi(\mu, \nu)
\end{pmatrix}.
\eea
The functions \( \phi \) and \( \psi \) fulfill these criteria,
\bea
\phi(0, 0) = 0, \quad \nabla \phi(0, 0) = 0\,, \nonumber\\ 
 \psi(0, 0) = 0, \quad \nabla \psi(0, 0) = 0\,, \nonumber
\eea 
with $\nabla$ denoting the gradient operator. In this context, $A$ and $B$ are square matrices that possess the following characteristics, 
\begin{itemize} 
\item The eigenvalues of $A$ have zero as their real parts. 
\item The eigenvalues of $B$ have negative real parts. 
\end{itemize}
  \item \textbf{Finding the Function $h(\mu)$:} Following that, a function $h(\mu)$ is determined, typically utilizing a series expansion. This function $h(\mu)$ fulfills the subsequent quasilinear partial differential equation,
  \bea \label{approximationfunction}
\mathcal{N}(h(\mu))=\nabla h(\mu)[A\mu+\phi(\mu,h(\mu))]-Bh(\mu)-\psi(\mu,h(\mu))=0 \,,
\eea
under the conditions $h(0) = 0$ and $\nabla h(0) = 0$.
   \item \textbf{Centre Manifold Dynamics:} By using the approximate solution of $h(\mu)$ derived from Eq.~\eqref{approximationfunction}, the behavior of the original system confined to the Centre manifold is expressed as follows,
   \bea \label{dynamicsofcentermanifoldequation}
   \mu' = A\mu + \phi(\mu, h(\mu))\,,
   \eea
   for $\mu \in \mathbb{R}$ is sufficiently small.
   \item \textbf{Final State of the Reduced System:} The equation $\mu' = A\mu + \phi(\mu, h(\mu))$ can be simplified to the expression $\mu' = k \mu^n$, where $k$ stands for a constant and $n$ indicates a positive integer, specifically referring to the term with the smallest order in the series expansion.
   \begin{itemize}
       \item If $k<0$ and $n$ is an odd integer, it can be concluded that the system is stable, suggesting that the original system is also stable.
       \item In any other situation, the reduced and original systems will exhibit instability.
   \end{itemize}
\end{enumerate}

The analysis reveals that the critical points \(g\) and \(i^{\pm}\) are non-hyperbolic, as their associated eigenvalues include both negative real parts and zero values. This non-hyperbolicity renders linear stability analysis inconclusive for assessing their stability properties. Consequently, to obtain a deeper understanding of the local dynamics around these points, we proceed with a detailed investigation using the CMT in the subsequent section.

\subsection{Stability of non-hyperbolic critical point}
We will apply CMT to the system defined by Eqs.~\eqref{Auto_Sys} to evaluate the stability of the critical point $g$. The Jacobian matrix corresponding to the autonomous system \eqref{Auto_Sys} at the critical point $g$ is expressed as
\[
J(g) = 
\begin{bmatrix}
 -3 & \sqrt{6} \beta  & \sqrt{\frac{3}{2}} \beta  & \sqrt{\frac{3}{2}} & 0 \\
 0 & -3 & \frac{1}{2} & 0 & 0 \\
 0 & 0 & -4 & 0 & 0 \\
 0 & 0 & 0 & 0 & 0 \\
 -\frac{2 y_c}{\sqrt{\alpha }} & 0 & 0 & 0 & 0 
\end{bmatrix}
\]

The eigenvalues of the  matrix $J(g)$ are $0,0,-4,-3,$ and $-3$.  The vectors \(\left[0, 0, 0, 0, 1 \right]^T\) and \(\left[-\frac{\sqrt{\alpha}}{2y_{c}}, 0, 0, -\frac{\sqrt{3 \alpha}}{\sqrt{2} y_{c}}, 1 \right]^T\) are eigenvectors corresponding to the eigenvalue \(0\), while \(\left[0,1,-2,0,0\right]^T\) is the eigenvector associated with the eigenvalue \(-4\). Similarly, the vectors \(\left[1,0,0,0,\frac{2 y_{c}}{3 \sqrt{\alpha}} \right]^T\) and \(\left[ 1, \frac{1}{\sqrt{6}\beta}, 0, 0, \frac{8y_c}{6\sqrt{\alpha}}\right]^T\) correspond to the eigenvalue \(-3\). To proceed, we need to transform this system into the format of \eqref{ReformulationoftheDynamicalSystem}. To shift the point \((0, 1, 0, 0, y_{c})\) to the origin of the phase space, we will introduce new coordinates: \( U=X \), \( V=Y-1 \), \( Q=\xi \), \( R=\lambda \), and \( S=y- y_{c} \). The system of equations is subsequently reformulated in terms of these new coordinates as
\begin{align}
\begin{pmatrix}
U'\\ 
V' \\ 
Q'\\ 
R'\\ 
S' 
\end{pmatrix}= 
\begin{pmatrix}
-3 & 0 & 0 & 0 & 0\\
0 & -3 & 0 & 0 & 0\\
0 & 0 & -4 & 0 & 0 \\
0 & 0 & 0 & 0 & 0\\
0 & 0 & 0 & 0 & 0
\end{pmatrix} 
\begin{pmatrix}
U\\ 
 V\\ 
 Q\\ 
 R\\ 
 S    
\end{pmatrix}+\begin{pmatrix}
 non\\ linear\\ term   
\end{pmatrix}. 
\end{align}

About the general formulation outlined in \eqref{ReformulationoftheDynamicalSystem}, we identify $U$, $V$, and $Q$ as the stable variables, while $R$ and $S$ serve as the central variables in the new system. According to the CMT, a continuous differentiable function characterizes the manifold. In this context, we have defined the Centre manifold as \( U = h_1(R, S) \), \( V = h_2(R, S) \), and \( Q = h_3(R, S) \).
\begin{eqnarray}
 U'= \frac{\partial h_1}{\partial R} R' + \frac{\partial h_1}{\partial S} S', \\ \label{CMT4}
 V'= \frac{\partial h_2}{\partial R} R' + \frac{\partial h_2}{\partial S} S',\\ \label{CMT5}
 Q'= \frac{\partial h_3}{\partial R} R' + \frac{\partial h_3}{\partial S} S'. \label{CMT6}
\end{eqnarray} 

Utilizing the framework given in Eq.~\eqref{approximationfunction}, we now construct the corresponding approximation as
\begin{eqnarray}
\mathcal{N}_{1}(h_1(R,S)) &=& h_1(R,S)_{,R}R'+   h_1(R,S)_{,S}S'-U'\,, \label{CMT7} \\ 
\mathcal{N}_{1}(h_2(R,S))&=& h_2(R,S)_{,R}R' + h_2(R,S)_{,S}S'-V'\,,\\ \label{CMT8}
\mathcal{N}_{1}(h_3(R,S)) &=& h_3(R,S)_{,R}R'+ h_3(R,S)_{,S}S'-Q'\,. \label{CMT9}
\end{eqnarray}

Where $h_1(R,S)_{,R}$, $h_2(R,S)_{,R}$ and $h_3(R,S)_{,R}$ denote the derivatives with respect to $R$, while $h_1(R,S)_{,S}$, $h_2(R,S)_{,S}$ and $h_3(R,S)_{,S}$ represented derivative with respect to $S$. After substituting the transformation into the system \eqref{Auto_Sys}, we obtain the following set of equations expressed in the new coordinate system,
 \begin{eqnarray}
U'&=& \frac{1}{2} \left(3 U^3+\sqrt{6} \beta  U^2+U \left(Q -3 (V+1)^2-3\right) \right.\nonumber\\
 && \left. +\sqrt{6} \left(\beta  \left(Q +(V+1)^2-1\right)+R  (V+1)^2\right)\right),\nonumber\\
 V'&=& \frac{1}{2} (V+1) \left(Q +3 U^2-\sqrt{6} R  U-3 (V+1)^2+3\right),\nonumber\\
 Q'&=& Q  \left(Q +3 U^2-3 (V+1)^2-1\right),\nonumber\\
 R'&=&\frac{\sqrt{\frac{3}{2}} R^2 U \left((S+y_{c})^2 (p-n)+2 n (S+y_{c})-n+p\right)}{(S+y_{c}) (n (S+y_{c})-n+p)^2},\nonumber\\
  S'&=& -\frac{2 U (S+y_{c})}{\sqrt{\alpha }}.
 \end{eqnarray}\label{CMT4-2} 

For zeroth approximation, we obtain the following approximation from Eqs.~(\ref{CMT7}-\ref{CMT9}), 
\begin{eqnarray}
\mathcal{N}_{1}(h_1(R,S)) &=& \sqrt{\frac{3}{2}}R\,, \label{CMT71} \\ 
\mathcal{N}_{1}(h_2(R,S))&=&0 \,,\\ \label{CMT81}
\mathcal{N}_{1}(h_3(R,S)) &=&0  \,. \label{CMT91}
\end{eqnarray}

Based on the zeroth-order approximation functions given in Eqs.~(\ref{CMT71}--\ref{CMT91}), the central variables are obtained as,
\begin{eqnarray}
R'&=& \frac{3 R^3}{2 \left(y_c+S\right) \left(n y_c+n (S-1)+p\right)^2} \nonumber \\ && \left[n(-1+y_c -y_c^2)+p(1+y_c^2) \right] \nonumber \\ &&+\text{higher} \hspace{0.1cm} \text{order} \hspace{0.1cm} \text{term},\label{CMT10} \\
S'&=& -\frac{\sqrt{6} y_c }{\sqrt{\alpha}}R+ \text{higher} \hspace{0.1cm} \text{order} \hspace{0.1cm} \text{term}.\label{CMT11}
\end{eqnarray}

According to the behavior of the system in Eqs.~(\ref{CMT10}, \ref{CMT11}), CMT indicates that the critical point $g$ exhibits stable dynamics under the following parameter conditions: $n\in \mathbb{R}\,\,\land\,\, y_c>0\,\,\land\,\, p<\frac{3 n y_c^2-3 n y_c+3 n}{2 y_c^2+2}$. 

We will now evaluate the stability of the non-hyperbolic critical points \( i^{\pm} \). It is important to note that the eigenvalues at the critical point \( g \) are the same as those at the points \( i^{\pm} \), along with their respective coordinates, except for the coordinate \( y_c \). For the critical points \( i^{\pm} \), we have \( y_c = \frac{n \pm \sqrt{2np - p^2}}{n - p} \). Consequently, the mathematical calculation for the CMT regarding \( i^{\pm} \) remains unchanged, except that we replace \( y_c \) with \( \frac{n \pm \sqrt{2np - p^2}}{n - p} \). From Eqs.~(\ref{CMT10}, \ref{CMT11}), we proceed to derive the central variables for the critical points \( i^{\pm} \) as, 
{\smaller
\begin{eqnarray}
R'&=& \frac{3 R^3}{2 \left(\frac{n \pm \sqrt{2np - p^2}}{n - p}+S\right) \left(n\,\, \frac{n \pm \sqrt{2np - p^2}}{n - p}+n (S-1)+p\right)^2} \nonumber \\ && \bigg[n\bigg(-1+\frac{n \pm \sqrt{2np - p^2}}{n - p} -\bigg(\frac{n \pm \sqrt{2np - p^2}}{n - p}\bigg)^2\bigg)+\nonumber \\ && p\bigg(1+\bigg(\frac{n \pm \sqrt{2np - p^2}}{n - p}\bigg)^2\bigg) \bigg] \nonumber \\ &&+\text{higher} \hspace{0.1cm} \text{order} \hspace{0.1cm} \text{term},\label{CMT12} \\  
S'&=& -\sqrt{\frac{6}{\alpha}} \bigg(\frac{n \pm \sqrt{2np - p^2}}{n - p}\bigg) \,R+ \text{higher} \hspace{0.1cm} \text{order} \hspace{0.1cm} \text{term}.\label{CMT13}
\end{eqnarray}
}
According to CMT, the critical point $i^{\pm}$ demonstrates stable dynamics when the parameters satisfy the conditions: $n>0\land 0< p<n$. Outside of these parameter constraints, the critical point exhibits unstable behavior. 
 \section{Cosmological Perturbations}\label{perturbation}
This section delves into the evolution of linear scalar perturbations within the theory delineated in Eq.~\eqref{alphaatractor_vX}. We analyze these perturbations against the FLRW metric, focusing on the perturbed line element represented as follows \cite{Bardeen:1980kt},
 \bea
ds^2=-(1+2\alpha)dt^2+2\partial_{i}{\chi}dtdx^{i}+ a(t)^2\left[(1+2\zeta)\delta_{i j}+2 \partial_{i}\partial_{j}{E}\right]dx^{i}dx^{j},
 \eea 
 
where the quantities $\alpha$, $\chi$, $\zeta$ and $E$ are scalar perturbations, which depend on both $t$ and spatial coordinates $x^{i}$. Here, we use the notation $\partial_{i}{\chi}=\frac{\partial \chi}{\partial {x^{i}} }$. The scalar field $\phi(t)$ is decomposed as follows 
 \be
\phi=\bar{\phi}(t)+\delta{\phi}(t,x^{i})\,,
 \ee
where $\bar{\phi}(t)$ and $\delta{\phi}$ indicate the background scalar field and its perturbation, respectively. To move forward, we introduce the following gauge-invariant quantities \cite{Amendola:2020ldb} 
\bea
&& \delta{\phi}_{N}=\delta{\phi}+\dot{\phi}\left(\chi-a^2 \dot{E}\right), \:\:\: \delta{\rho}_{IN}=\delta{\rho}_{I}+\dot{\rho}_I\left(\chi-a^2 \dot{E}\right)\,, \nonumber\\
&& \Psi=\alpha+\frac{d}{dt}\left(\chi-a^2 \dot{E}\right), \:\:\: \Phi= \zeta+H\left(\chi-a^2 \dot{E}\right)\,, \nonumber\\
&& \nu_{I\,N}= \nu_{I}+\chi-a^2 \dot{E}\,,
\eea in conjunction with the dimensionless parameters,
\bea
&&\delta_{IN}=\frac{\delta \rho_{IN}}{\rho_I}, \:\:\:\:\:\: \delta \varphi_{N}=\frac{H}{\dot{\phi}}\delta \phi_{N},\nonumber\\ 
&&V_{IN}=H \nu_{IN}, \:\:\:\:\:\:  \mathcal{K}=\frac{k}{aH}\,,
\eea
where \( k \) denotes the comoving wavenumber and \( \nu_{I} \) represents the scalar velocity potential. Therefore, in Fourier space, the perturbed field equations, in their gauge-ready form, are expressed as \cite{Ryotaro_2020, Amendola:2020ldb}
\bea
&& 6 X^2\delta{\varphi}'_{N}-6 \Phi'+6 (1-X^2)\left(\hat{\xi}\delta{\varphi}_{N}+\Psi\right)-2\mathcal{K}^2 \Phi +\nonumber\\
&& 3 \left(3\Omega_{m}+4\Omega_r\right)\delta{\varphi}_{N}+3\left(\Omega_{m}\delta_{mN}+\Omega_r\delta_{rN}\right)=0, \label{pertur1} \\ 
&& \Phi'-\Psi-\hat{\xi}\delta{\varphi}_{N}+\frac{3}{2}\Omega_{m}\left(V_{m N}-\delta{\varphi}_{N}\right)\nonumber\\
&&+2\Omega_{r}\left(V_{rN}-\delta{\varphi}_{N}\right) =0, \label{pertur2}\\
&& \delta'_{IN} + 3 (c_{I}^{2}-\omega_{I}) \delta_{IN}+ (1+\omega_{I}) (\mathcal{K}^2 V_{IN}+ 3 \Phi')=0, \label{pertur3}\\
&& \Omega_{m} V'_{mN}-\left(\hat{\xi} \Omega_{m}-\sqrt{6}\beta X \Omega_{m}\right) V_{mN}-\Omega_{m}\Psi \nonumber\\&& - \sqrt{6} \beta X \Omega_{m}\delta{\varphi}_{N} =0, \label{pertur4}\\
&& V'_{IN}-(\hat{\xi} + 3 c_{I}^{2})V_{IN}- \Psi- \frac{c_I^2}{1+\omega_I}\delta_{IN}=0, \label{pertur5}\\
&& \delta{\varphi}''_{N}+ (3-\hat{\xi}+2 \epsilon_{\phi})\delta{\varphi}'_{N}+ \bigg[\mathcal{K}^2-\hat{\xi}'-3 \hat{\xi}+ \epsilon_{\phi}'+\epsilon_{\phi}^2 \nonumber\\&&+(3-\hat{\xi})\epsilon_{\phi} +\hat{M}_\phi^2+3\beta^2 \Omega_{m} \bigg]\delta{\varphi}_{N}+ 3 \Phi' - \Psi' \nonumber\\&&- 2(3+\epsilon_{\phi})\Psi+ \frac{\sqrt{6} \beta \Omega_{m}}{2 X} \delta_{mN}=0,\label{pertur6}\\
&& \Psi = -\Phi,\label{pertur7}
\eea 

where $\hat{\xi}=\frac{H'}{H}$ and $\epsilon_{\phi}=\frac{\ddot{\phi}}{\dot{\phi}H}$ . The quantities $c_I^2$ and $\omega_I$ represent each matter component's sound speed and the EoS parameter, where $I = m\,, r$ corresponds to matter and radiation, respectively. The prime ($'$) represents a derivative with respect to $N$. In Eq.~\eqref{pertur6}, we have defined
{\smaller
\bea
&&\hat{M}_\phi^2= \frac{M_{\phi}^2}{H^2}\equiv \frac{1}{H^2} \frac{d^2V(\phi)}{d^2\phi}=3\Gamma \lambda^2 Y^2 \nonumber\\
&&=\frac{3 \lambda ^2 Y^2}{2 \left(\frac{p-n}{y}+n\right)^2}\Bigg[2 n^2+\frac{2 \left(p^2+(n-1) n-2 n p\right)}{y^2}+\frac{n-p}{y^3}- \frac{(4 n-1) (n-p)}{y}\Bigg]\,,
\label{dimensionlessmass} 
\eea } 

where the function $M_{\phi}^2$ denotes the field mass squared associated with the scalar potential in Eq. \eqref{spot} \cite{Kase:2019veo}. The variable $\Gamma = V V_{,\phi\phi}/(V_{,\phi})^2$ encodes the concavity of the potential. In our analysis it is not treated as an independent variable. Once the potential is specified, $\Gamma$ can be expressed in terms of the other dynamical variables and substituted directly into the equations. Its explicit role appears only in Eq.~(5.12), where it determines the 
effective scalar mass squared, $m_{\text{eff}}^2 = V_{,\phi\phi} = (V_{,\phi})^2 \Gamma / V$. The effective mass governs the behaviour of scalar-field fluctuations. For a light field, $m_{\text{eff}}^2 \ll H^2$, perturbations can evolve and cluster, thereby affecting the growth of structure. For a heavy field, $m_{\text{eff}}^2 \gg H^2$, perturbations are suppressed and the field remains essentially homogeneous \cite{DeFelice:2010aj,amendola2010dark,Ryotaro_2020}.

We now examine the evolution of matter perturbations. For this purpose, we ignore the contribution of radiation perturbation in Eqs.~(\ref{pertur1}-\ref{pertur7}) by setting $\Omega_{r} = 0$. Under this assumption, the matter sector simplifies, and Eq.~\eqref{pertur3} takes the following form
\begin{equation}\label{matterperturbation}
\delta'_{mN}+ \mathcal{K}^2 V'_{mN}+ 3 \Phi'=0.  
\end{equation}

\begin{figure}[!t]
    \centering
        \includegraphics[scale=0.45]{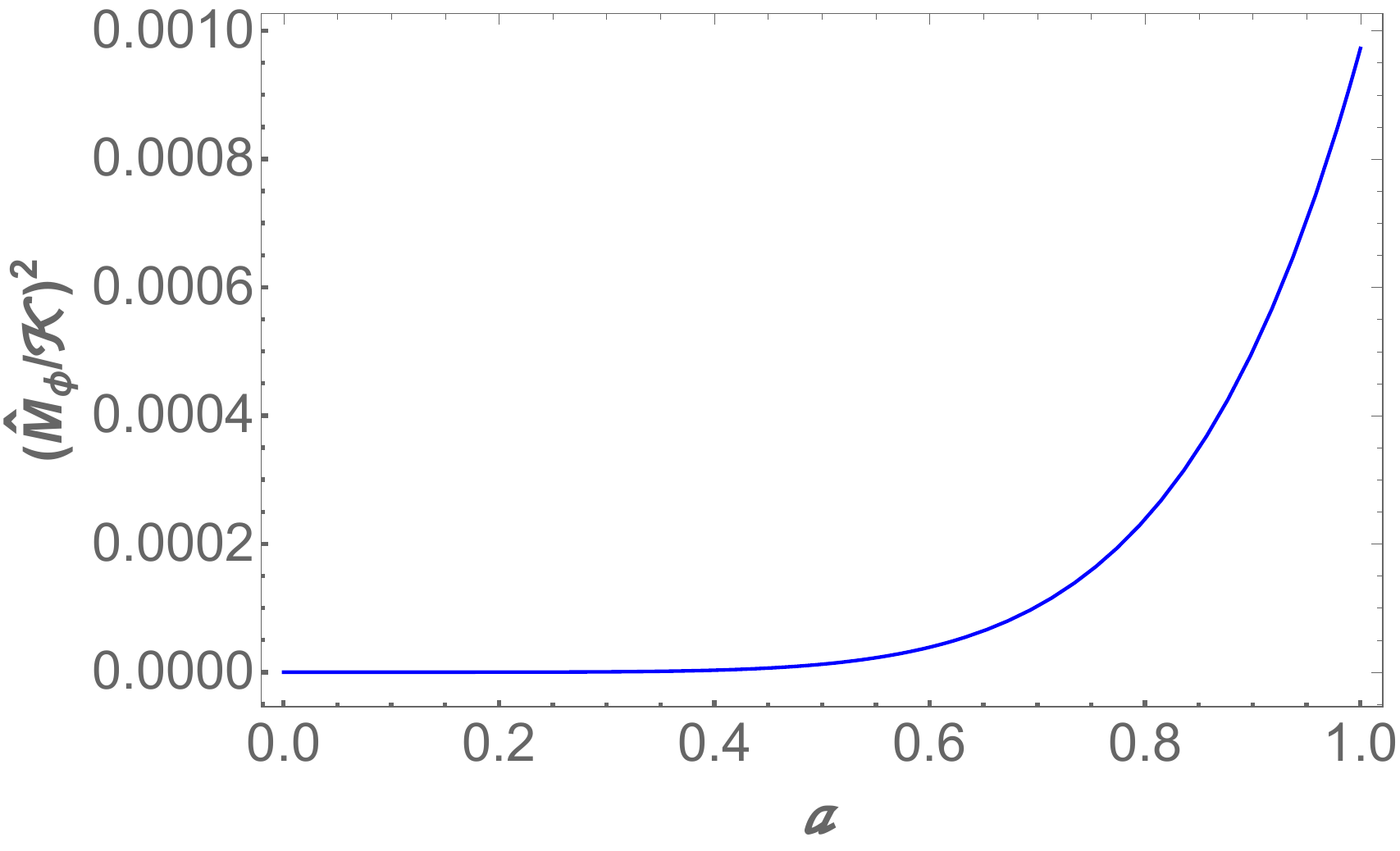}
        \caption{Evolution of the field mass squared. We have assumed the linear scale $k = 0.1~h~\text{Mpc}^{-1}$ and parameter values $n = 2$, $p = 1$, $\alpha = 1$, using the same initial conditions as in Figure \ref{numerical_phase}. }
    \label{FIG_Mphi2}
\end{figure}

In accordance with Figure~\ref{FIG_Mphi2}, we neglect $\hat{M}_{\phi}^2$ in comparison to $\mathcal{K}^2$.
Since dark energy does not cluster on astrophysical scales, it does not exhibit the behavior characteristic of massive dark matter \cite{Matos_2000, Amendola_2004prd}. Under this assumption, we differentiate Eq.~\eqref{matterperturbation} with respect to $N$, and using the result from Eq.~\eqref{pertur4}, we obtain the following expression
\begin{equation}\label{matterper1}
\delta''_{mN}+\mu_1 \delta'_{mN}+ \mu_2 \mathcal{K}^2 + 3 \Phi'' + 3 \mu_1 \Phi'=0,     
\end{equation}
where 
\begin{eqnarray} \label{mu1mu2}
&&\mu_1 = 2+ \hat{\xi}+\sqrt{6} \beta X, \nonumber\\
&&\hspace{0.2cm} \mu_2 = \Psi + \sqrt{6} \beta X \delta{\varphi}_{N}.    
\end{eqnarray}

Now, we employ the quasi-static approximation for perturbations deep inside the sound horizon \cite{De_Felice_2011, Esposito_Far_se_2001, Amendola_2020jcap}. The dominant contributions to the perturbation equations arise from $\mathcal{K}^2$, $\delta_{mN}$, and $\delta'_{mN}$, Eqs.~\eqref{pertur1}, \eqref{pertur6}, and \eqref{pertur7} yield the following relationships,
\begin{eqnarray}
 \Psi = -\Phi \approx -\frac{3}{2 \mathcal{K}^2} \Omega_{m} \delta_{mN}, \label{approx1}\\
\delta{\varphi}_{N} \approx -\frac{\sqrt{6} \beta \Omega_{m} \delta_{mN}}{2 X \mathcal{K}^2}.\label{approx2}    
\end{eqnarray}

\begin{figure}[h]
    \centering
        \includegraphics[scale=0.42]{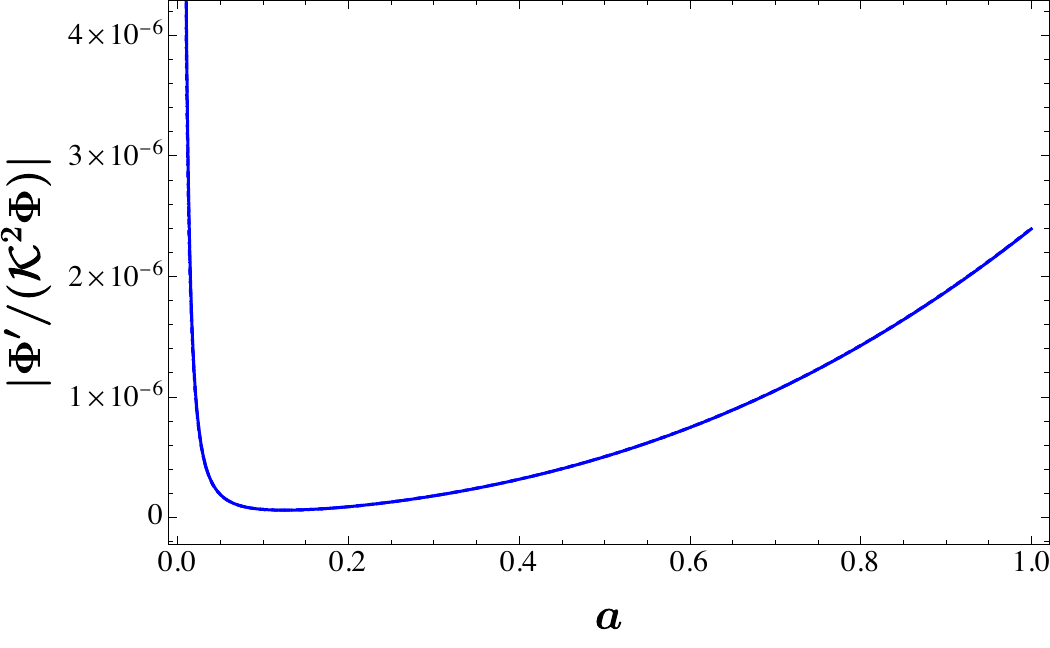} \
        \includegraphics[scale=0.42]{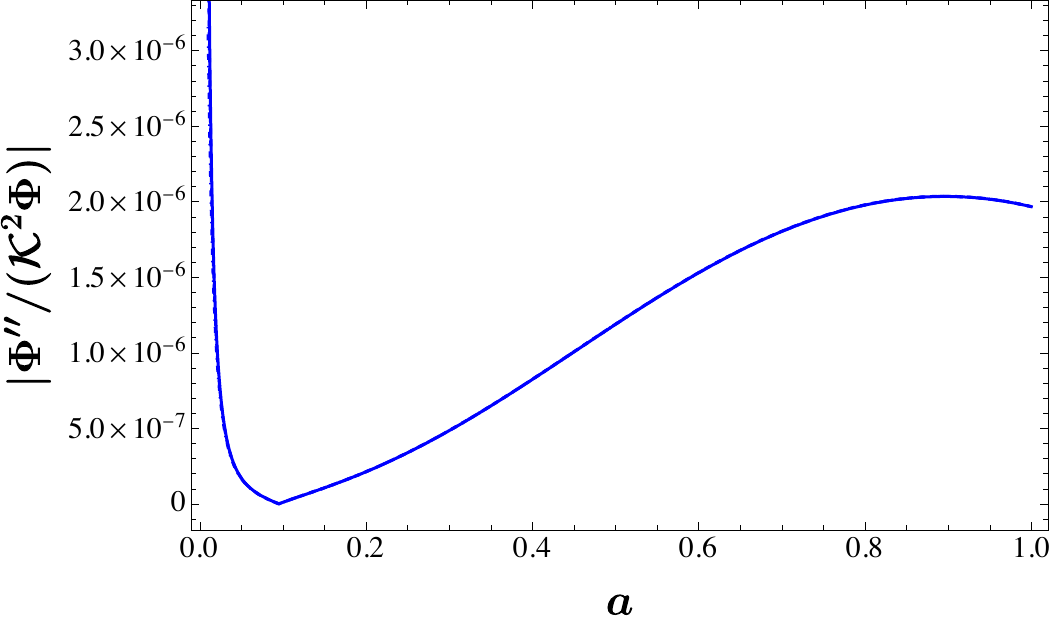}
        \includegraphics[scale=0.42]{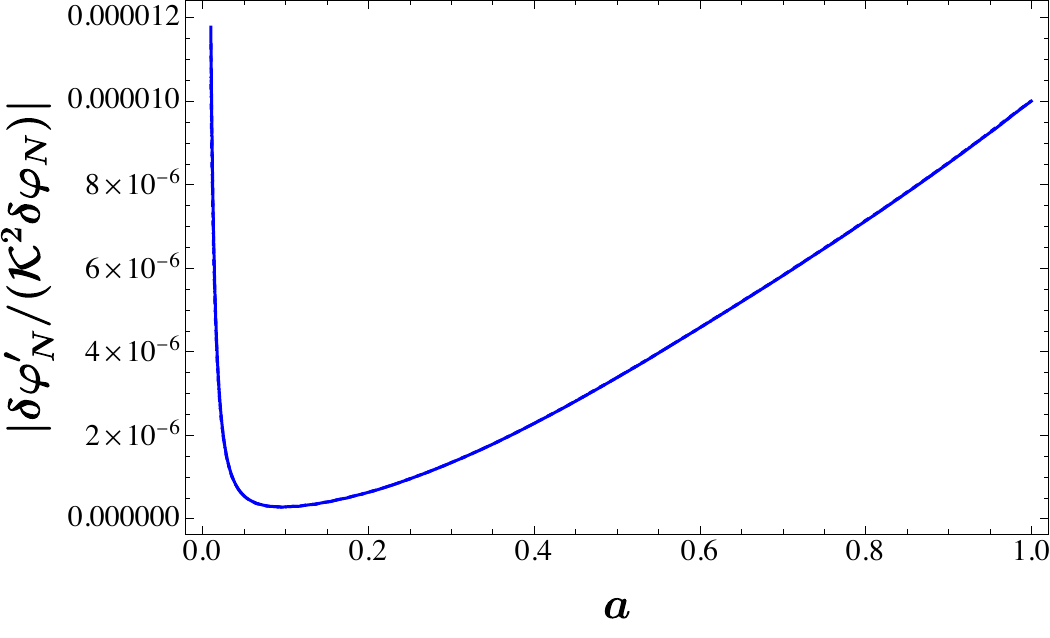}
        \caption{Top left panel: Time–derivative terms of the metric potentials, $\left|\frac{\Phi'}{\mathcal{K}^2 \Phi}\right| \ll 1$. Top right panel: Time–derivative terms of the metric potentials, $\left|\frac{\Phi''}{\mathcal{K}^2 \Phi}\right| \ll 1$.  Botton panel: Time–derivative terms of scalar perturbations, $\left|\frac{\delta\varphi_N'}{\mathcal{K}^2 \delta\varphi_N}\right| \ll 1$. We have assumed the linear scale $k = 0.1~\text{Mpc}^{-1}$ and parameter values $n = 2$, $p = 1$, $\alpha = 1$, using the same initial conditions as in Figure \ref{numerical_phase}.}  
    \label{fig:potencials_time}
\end{figure}

In the context of the quasi-static approximation, we can neglect the terms \(3\Phi''\) and  \(3\mu_1\Phi'\) in Eq.~\eqref{matterper1} relative to the remaining terms, as confirmed in Fig.~\ref{fig:potencials_time}, where the ratios \( |\Phi'/(\mathcal{K}^2\Phi)| \), \( |\Phi''/(\mathcal{K}^2\Phi)| \), and \( |\delta\varphi_N'/(\mathcal{K}^2\delta\varphi_N)| \) remain \(\ll 1\) for \(\mathcal{K}\gg1\). By substituting Eqs.~(\ref{approx1}, \ref{approx2}) and the \(N\) derivative of Eq.~\eqref{approx2} into Eq. \eqref{matterper1}, we derive the matter density contrast evolution equation
\begin{equation} 
    \delta''_{mN}+\bigg(2+ \frac{H'}{H}+ \sqrt{6} \beta X\bigg) \delta'_{mN}-\frac{3}{2G}G_{mm}\Omega_{m}\delta_{mN} = 0.
    \label{MatterPert}
\end{equation}

The relation between effective gravitational constant $G_{mm}$ and gravitational constant $G$ is 
\begin{equation}\label{GmmG}
G_{mm}= (1+2 \beta^2) G\,.     
\end{equation}
For $\beta=0$, the effective gravitational constant $G_{mm}$ reduces to the standard gravitational constant $G$. When $\beta \neq0$, $G_{mm}$ exceeds $G$.

\subsection{Analytical results}

By introducing the growth rate of matter perturbations $f_{\delta}\equiv \delta'_{mN}/\delta_{mN}$, Eq. \eqref{MatterPert} can be rewritten as 
\be
f_{\delta}'+f_{\delta}^2+\left(2+\frac{H'}{H}+\sqrt{6}\beta X\right)f_{\delta}-\frac{3}{2G}G_{mm}\Omega_{m}=0, 
\label{GrowthRate}
\ee where $H'/H= -1-q$. On the critical points, $f_{\delta}$ and the coefficients in Eq. \eqref{GrowthRate} are constants. Thus, we obtain 
\bea
 f_{\delta}=-\frac{1}{2}\Bigg[-1-q+\sqrt{6}\beta X+ \sqrt{\left(-1-q+\sqrt{6}\beta X\right)^2+\frac{6 G_{mm}}{G}\Omega_{m}}\Bigg].
\eea
For the matter scaling solution $f$ we get
\be
f_{\delta}\approx 1+2 \beta^2,
\label{growrate_scaling}
\ee
and thus
\be
\delta_{mN}(a)\propto a^{1+2\beta^2}, 
\ee in agreement with the results shown in Ref. \cite{Amendola_2020jcap}. 

The observable $f\sigma_8(z)$ compares model predictions with observations and is defined as
\be
f\sigma_{8}(z)\equiv f_{\delta}(a)\cdot \sigma(a)=\frac{\sigma_{8}}{\delta{(1)}} a \delta'(a)\label{fsigma8_eq}, 
\ee where $\sigma(a)=\sigma_{8} \delta(a)/\delta(1)$ is the amplitude of matter density fluctuations
in spheres of size $8~h^{-1}~\text{Mpc}$ ($k\sim k_{\sigma_{8}}=0.125$~$h~\text{Mpc}^{-1}$) with $\sigma_{8}=\sigma(1)$ strongly influenced by late-time expansion and dark energy models \cite{Sola:2017znb,Gonzalez-Espinoza:2018gyl}. Tensions in the 
$\Lambda$CDM model, particularly regarding 
$H_{0}$ and $\sigma_8$, arise as LSS data show $f\sigma_{8}$ values about $8\%$ lower than expected. This suggests that 
$\Lambda$CDM overestimates $\sigma_8$	
  for the same present-day growth rate $f_{\delta}(1)$ \cite{Gomez-Valent:2018nib}. However, since $f_{\delta}(a)$ is model-dependent \cite{DeFelice:2010aj}, a lower growth rate could also improve agreement with LSS data. To compare with $\Lambda$CDM model, we also define the growth index through the relation $f_{\delta}=\Omega_{m}^{\gamma}$ \cite{Wang:1998gt}. In the $\Lambda$CDM scenario, this index takes the value $\gamma=6/11\approx 0.55$ \cite{Linder:2005in,Nesseris:2007pa,Buenobelloso:2011sja}.  

\subsection{Numerical results}

\begin{figure}[!t]
    \centering
        \includegraphics[scale=0.5]{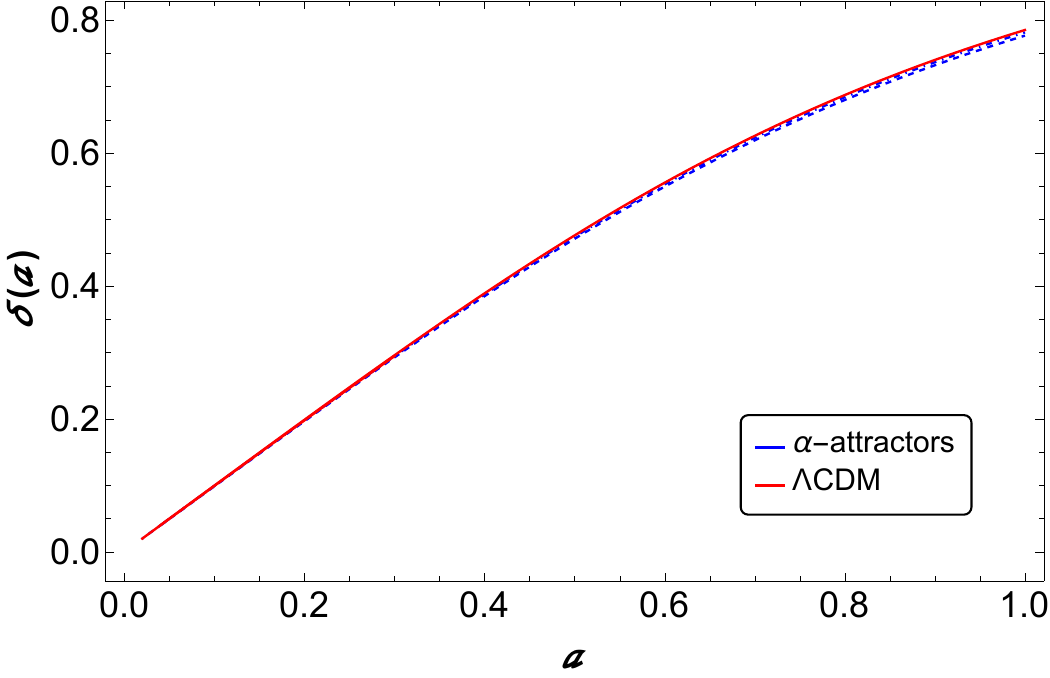}
        \caption{The graph illustrates the evolution of the matter density perturbation, denoted as $\delta(a)$, as a function of the scale factor $a$ for different values of the parameter $\beta$, the same conditions as those considered in Figure~\ref{numerical_phase}.
}
    \label{plotfdelta}
\end{figure}

Figure~\ref{plotfdelta} illustrates the theoretical curves for matter density perturbations with fixed parameters: $\alpha = 1$, $n = 2$, and $p = 1$. The variable $\beta$ is varied for three specific values: $\beta = 5.0 \times 10^{-3}$ (represented by the dot-dashed line), $\beta = 1.0 \times 10^{-2}$ (dashed line), and $\beta = 0$ (solid line). The initial conditions are set as follows: $X_0 = 10^{-11}$, $Y_0 = 8.6 \times 10^{-13}$, $\xi_0 = 0.999656$, $\lambda_0 = 1.0 \times 10^{-10}$, and $y_0 = 1.0 \times 10^{-11}$. For our analysis, we note that $\sigma(a) \sim \delta(a)$ \cite{Kazantzidis:2018rnb,Gonzalez-Espinoza:2018gyl}. From Figure~\ref{plotfdelta}, we estimate the following values: for $\beta = 5.0 \times 10^{-3}$ (dot-dashed line) and $\beta = 1.0 \times 10^{-2}$ (dashed line), the corresponding values of $\sigma_8 \sim \delta{(1)}$ are $\sigma_8 = 0.781$ and $\sigma_8 = 0.777$, respectively. In comparison, for the $\Lambda$CDM model, we find $\sigma_8 = 0.785$.

\begin{figure}[!t]
    \centering
        \includegraphics[scale=0.5]{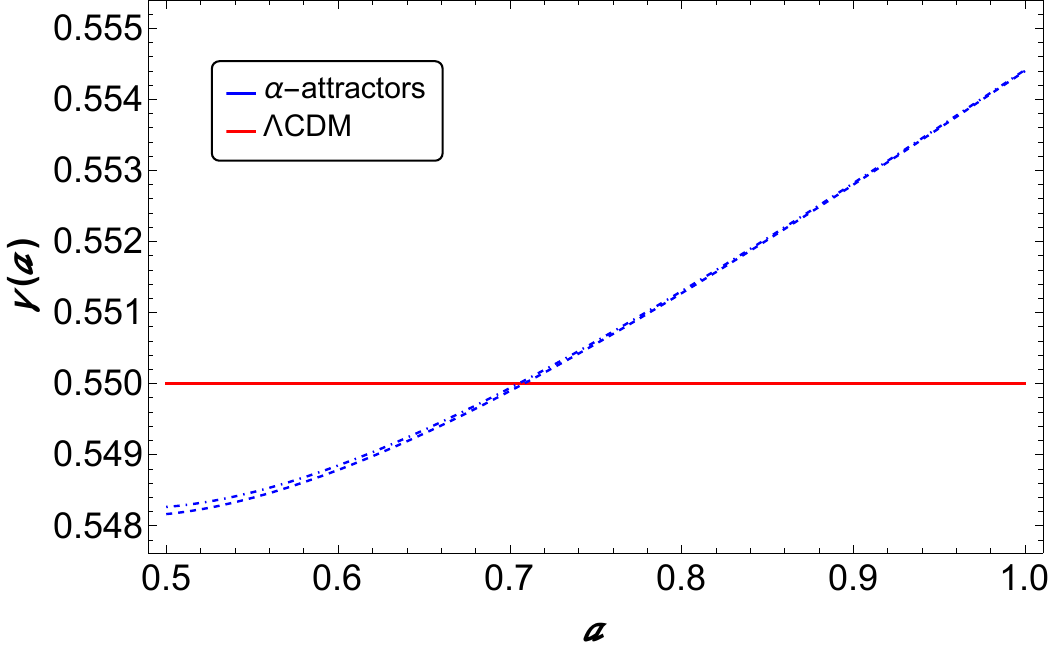}
        \caption{This graph illustrates the behavior of the growth index \(\gamma(a)\). We observe a deviation from the standard value of \(\gamma = 0.55\) for the \(\Lambda\)CDM model, with \(\gamma(a)\) showing larger values. This suggests that the growth rate of matter perturbations is smaller for the different parameter values of \(\beta\) used in Figure~\ref{numerical_phase}.
}
    \label{plotgamma}
\end{figure}

We compute the growth index using the relation $\gamma = \frac{\log{f_{\delta}}}{\log{\Omega_{m}}}$. Its behavior for different values of $\beta$ at the linear scale $k = 0.1~h~\text{Mpc}^{-1}$ is shown in Figure \ref{plotgamma}. During the scaling matter regime, before the domination of dark energy, the analytical solution for the growth rate is greater than one, according to Eq.~\eqref{growrate_scaling}, due to the dark energy-dark matter interaction. This is consistent with the behavior observed for the growth index in Figure \ref{plotgamma}. In this regime, the growth index becomes smaller than the $\Lambda$CDM value, implying a suppression of the growth rate of matter linear perturbations. When dark energy starts to dominate, the growth rate is smaller than one because of the accelerated expansion ($q<0$), which implies that the growth index becomes greater than $\Lambda$CDM value. As dark energy begins to dominate, the growth rate drops below one due to the accelerated expansion, and the growth index subsequently becomes larger than the $\Lambda$CDM value, consistent with Figure \ref{plotgamma}.

\begin{figure}[!t]
    \centering
        \includegraphics[scale=0.5]{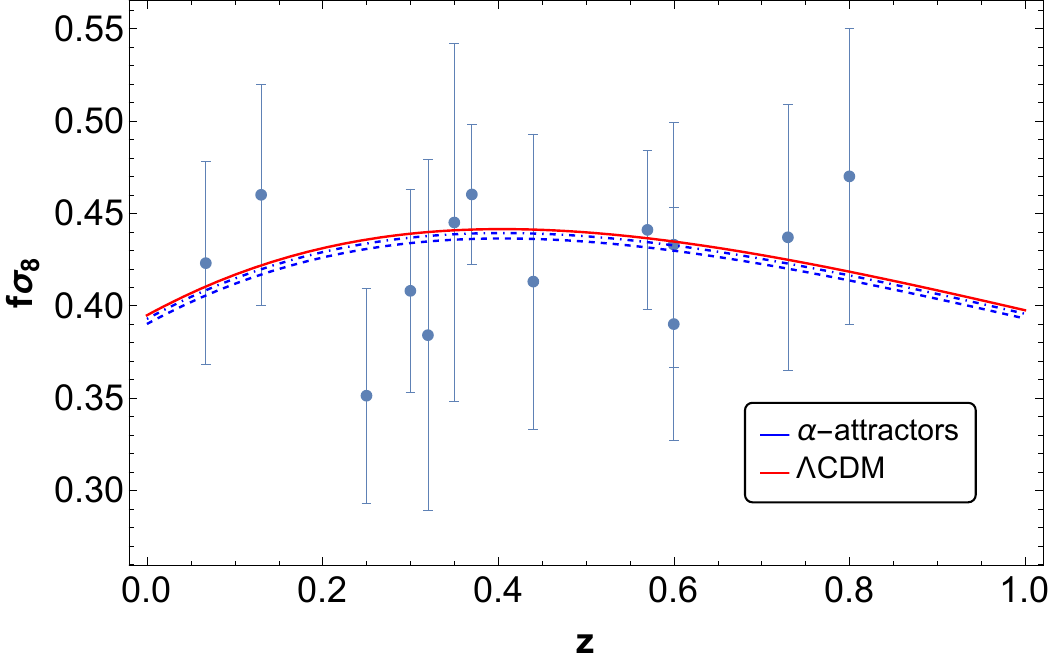}
        \caption{The evolution of the weighted growth rate $f\sigma_8$ is presented under the same conditions as those considered in Figure~\ref{numerical_phase}, along with the corresponding evolution for the $\Lambda$CDM model. In this plot, we have utilized the full redshift-space distortion (RSD) $f\sigma_8$ dataset from Table II in Ref.~\cite{Kazantzidis:2018rnb}.
}
    \label{plotfS8}
\end{figure}

Figure~\ref{plotfS8} presents the theoretical curves for the weighted linear growth rate \( f\sigma_8(z) \) with values of \( \beta = 5.0 \times 10^{-3} \) and \( \beta = 1.0 \times 10^{-2} \). We observe that these values are lower than the corresponding predictions from the \(\Lambda\)CDM model, indicating that our model may have the potential to alleviate the \(\sigma_8\)-tension.

To quantify this effect, we define the exact relative difference as
\begin{equation}
\Delta f\sigma_8 (z) \equiv 100 \times \frac{f\sigma_8(z)_{\text{model}} - f\sigma_8(z)_{\Lambda\text{CDM}}}{f\sigma_8(z)_{\Lambda\text{CDM}}},
\end{equation}
with respect to the concordance model \cite{Gomez-Valent:2018nib}. For $\beta=5.0 \times 10^{-3}$, we obtain $f\sigma_8(0) \approx 0.393$, while for $\Lambda$CDM, we find $f\sigma_8(0)_{\Lambda\text{CDM}} \approx 0.395$, leading to a relative difference of $\Delta f\sigma_8(0) \approx 1\%$. Consequently, the prediction for $\beta=5.0 \times 10^{-3}$ is approximately $1\%$ lower than that of $\Lambda$CDM. For $\beta=1.0 \times 10^{-2}$, we obtain $f\sigma_8(0) \approx 0.390$, leading to a relative difference of $\Delta f\sigma_8(0) \approx 2\%$, indicating a $2\%$ lower prediction than $\Lambda$CDM.


\subsubsection{Chi-square analysis with $f\sigma_8(z)$ data}
\begin{figure}[!h]
    \centering
        \includegraphics[scale=0.6]{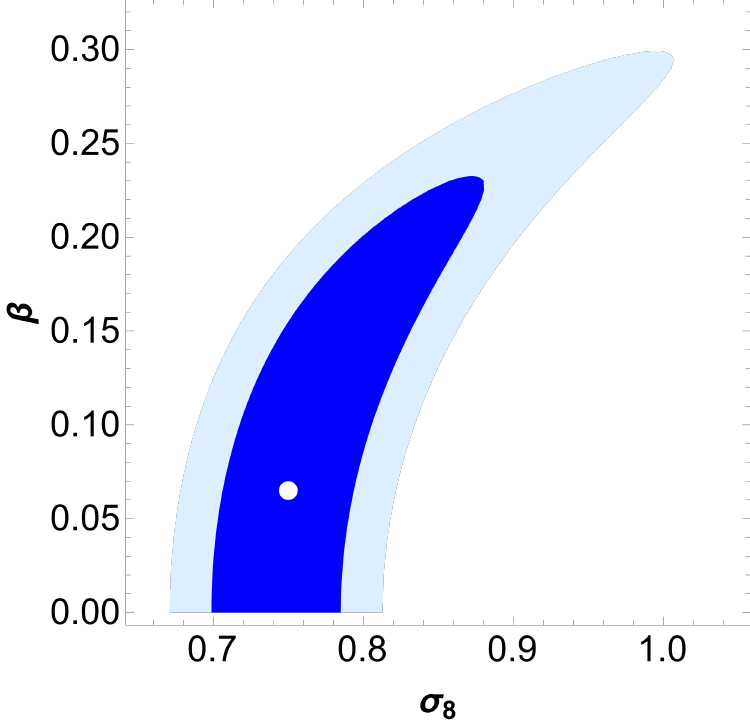}
        \caption{Constraints on parameters $\sigma_8$ and $\beta$. }
    \label{FIG_contour}
\end{figure}

We use the redshift-space distortion compilation of growth-rate measurements \( \mathbf{d}=\{f\sigma_8(z_i)\}_{i=1}^{N} \) with covariance \(C\), see Ref. \cite{Kazantzidis:2018rnb}. For parameters \( \boldsymbol{\theta} = \{\sigma_8, \beta \} \), the prediction is \( m_i(\boldsymbol{\theta}) = f\sigma_8(z_i;\boldsymbol{\theta})\, \), Eq. \eqref{fsigma8_eq}. The statistic and likelihood are
\[
\chi^2(\boldsymbol{\theta})=(\mathbf{d}-\mathbf{m})^{\!\top}C^{-1}(\mathbf{d}-\mathbf{m}).
\]
We minimize \(\chi^2\) to obtain \( \boldsymbol{\hat\theta} = \{ 0.750, 0.065 \} \) and report \( \chi^2_{\min} = 4.427\), the reduced value \( \chi^2_\nu=\chi^2_{\min}/(N-2) = 0.402 \), see Figure \ref{FIG_contour}.\\

For the best-fit parameters $\sigma_8=0.750$ $\beta=0.065$, we obtain $f\sigma_8(0) \approx 0.387$, leading to a relative difference of $\Delta f\sigma_8(0) \approx 2\%$, indicating a $2\%$ lower prediction than $\Lambda$CDM.

\section{Discussions and final Remarks}\label{conclusions}
In this work, we have explored the cosmological dynamics of $\alpha$-attractor models within the context of dark energy, incorporating an interaction between dark matter and dark energy. By performing a dynamical system analysis, we identified a range of critical points representing different cosmological eras, including radiation-dominated, matter-dominated, and accelerated expansion phases. Notably, we found scaling solutions where dark energy and dark matter evolve together, potentially addressing the cosmic coincidence problem.

Through numerical simulations, we analyzed the late-time behavior of the universe under the $\alpha$-attractor framework. Our results show that for a certain range of parameters, the model allows for a viable transition from radiation and matter domination to an accelerated expansion, consistent with observational constraints on the dark energy EoS parameter.

We also investigate the evolution of linear matter perturbations and their impact on the growth rate of cosmic structures. We derived and solved the evolution equation for $f\sigma_8$, comparing its predictions to $\Lambda$CDM. Our findings suggest that for specific choices of the interaction parameter, the model leads to a lower $f\sigma_8$, which could help alleviate the well-known $\sigma_8$ -tension observed in large-scale structure surveys \cite{DiValentino:2025sru}. This implies that the $\alpha$-attractor interacting model may provide a better fit to structure formation data compared to standard $\Lambda$CDM.

In future work, it would be interesting to further investigate the role of  $\alpha$-attractor models in addressing the $H_0$ tension, as these models may provide novel insights into late-time cosmic acceleration \cite{Garcia-Garcia:2018hlc,Dimopoulos:2023tcc,Braglia:2020bym}. Moreover, exploring the non-linear regime could shed light on structure formation and possible deviations from standard cosmology \cite{Euclid:2024jej,Baldi:2022uwb,Baldi:2010vv,Barreira:2013xea}. Such extensions would complement the present analysis and open new avenues for testing the robustness of the theoretical framework with observational data.

\begin{acknowledgments}
B.M acknowledges the support of Council of Scientific and Industrial Research (CSIR) for the project grant (No. 03/1493/23/EMR II). G.O acknowledges Dirección de Investigación, Postgrado y Transferencia Tecnológica de la Universidad de Tarapacá for financial support through Proyecto UTA Mayor 4742-25. M.G-E. acknowledges the financial support of FONDECYT de Postdoctorado, N° 3230801.
\end{acknowledgments}


\bibliographystyle{spphys}   
\bibliography{bio}

	

\end{document}